\documentclass[a4paper,fleqn,usenatbib]{mnras}

\usepackage{newtxtext,newtxmath}

\usepackage[T1]{fontenc}
\usepackage{ae,aecompl}
\usepackage{comment}

\usepackage{graphicx}	
\usepackage{amsmath}	
\usepackage{amssymb}	

\newcommand{\kms}{\,km\,s$^{-1}$} 
\newcommand{\cf}{{\ifmmode{C_{\rm f}}\else{$C_{\rm f}$}\fi}}
\newcommand{\zl}{{\ifmmode{z_{\rm l}}\else{$z_{\rm l}$}\fi}}
\newcommand{\zem}{{\ifmmode{z_{\rm em}}\else{$z_{\rm em}$}\fi}}
\newcommand{\zabs}{{\ifmmode{z_{\rm abs}}\else{$z_{\rm abs}$}\fi}}
\newcommand{\cm}{{\ifmmode{{\rm cm}^{-1}}\else{cm$^{-1}$}\fi}}
\newcommand{\cmm}{{\ifmmode{{\rm cm}^{-2}}\else{cm$^{-2}$}\fi}}
\newcommand{\CMM}{{\ifmmode{{\rm cm}^{2}}\else{cm$^{2}$}\fi}}
\newcommand{\cmmm}{{\ifmmode{{\rm cm}^{-3}}\else{cm$^{-3}$}\fi}}
\newcommand{\lya}{{\rm Ly}$\alpha$} 
\newcommand{\ew}{{\ifmmode{EW}\else{$EW$}\fi}}
\newcommand{\ews}{{\ifmmode{EW{\rm s}}\else{$EW{\rm s}$}\fi}}

\newcommand{\lam}{$\lambda$}
\newcommand{\kpc}{{\ifmmode{\rm kpc}\else{${\rm kpc}$}\fi}}
\newcommand{\pc}{{\ifmmode{\rm pc}\else{${\rm pc}$}\fi}}
\newcommand{\Mpc}{{\ifmmode{\rm Mpc}\else{${\rm Mpc}$}\fi}}
\newcommand{\K}{{\ifmmode{\rm K}\else{${\rm K}$}\fi}}

\usepackage{lscape}
\usepackage{longtable}
\usepackage{threeparttable}

\newcounter{species} 

\def\ion#1#2{\setcounter{species}{#2}#1$\;${\scriptsize\Roman{species}}\relax}
\title[Intrinsic NALs in BAL/mini-BAL quasar spectra]{Search for intrinsic NALs in BAL/mini-BAL quasar spectra}

\author[D. Itoh et al.]{
Daisuke Itoh,$^{1}$\thanks{E-mail: 18hs301e@shinshu-u.ac.jp (DI)}
Toru Misawa,$^{2}$
Takashi Horiuchi$^{3}$
and 
Kentaro Aoki$^{4}$
\\
$^{1}$Department of Physics, Faculty of Science, Shinshu University, 3-1-1 Asahi, Matsumoto, Nagano 390-8621, Japan\\
$^{2}$School of General Education, Shinshu University, 3-1-1 Asahi, Matsumoto, Nagano 390-8621, Japan\\
$^{3}$Ishigakijima Astronomical Observatory, National Astronomical Observatory of Japan, National Institutes of Natural Sciences, 1024-1 Arakawa, Ishigaki, \\ Okinawa, 907-0024, Japan\\
$^{4}$Subaru Telescope, National Astronomical Observatory ofJapan, 650 North A’ohoku Place, Hilo, HI 96720, USA\\
}
\date{Accepted XXX. Received YYY; in original form ZZZ}

\pubyear{2019}

\begin{document}
\label{firstpage}
\pagerange{\pageref{firstpage}--\pageref{lastpage}}
\maketitle

\begin{abstract}
Some fraction of narrow absorption lines are physically associated to
the quasar/host-galaxy materials (i.e., intrinsic NALs) like those of
BALs and mini-BALs. The relation between these three types of
absorption lines has not been understood yet, however one
interpretation is that these absorption features correspond to
different inclination angles.  In this study, we search for intrinsic
NALs in 11 BAL/mini-BAL quasar spectra retrieved from VLT/UVES public
archive, in order to test a possible relation of intrinsic NALs and
BALs/mini-BALs in the geometry models.  We use partial coverage
analysis to separate intrinsic NALs from ones which are associated to
cosmologically intervening materials like foreground galaxies and
intergalactic medium (i.e., intervening NALs).  We identify one
  reliable and two possible intrinsic NAL systems out of 36 NAL
  systems in 9 BAL/mini-BAL quasar spectra after removing two quasars
  without clear BAL features.  In spite of a small sample size, we
  placed a lower limit on the fraction of BAL/mini-BAL quasars that
  have at least one intrinsic \ion{C}{4} NAL
  ($\sim33^{+33}_{-18}\%$). This can be interpreted that intrinsic
NAL absorbers exist everywhere regardless of inclination angle.  We
found that one of the intrinsic NAL systems detected in SDSS
J121549.80-003432.1 is located at a large radial distance of
$R>130~{\kpc}$, using a method of photoionization model with
ground/excited-state lines.  Considering the wide range of intrinsic
NAL absorber distribution in inclination angles and radial distances,
it suggests that origins and geometry of them are more complicated
than we expected.

\end{abstract}




\section{Introduction}
\label{sec:intro}
A substantial fraction of absorption lines in quasar spectra are
physically associated to the background quasars (i.e., {\it intrinsic}
absorption lines), rather than the other absorbers like intergalactic
medium (IGM) or intervening galaxies ({i.e., \it intervening}
absorption lines).  They are caused by strong outflowing stream from
accretion disk (i.e., outflow winds), and their offset velocities from
the quasars are sometimes relativistic up to $\sim0.3c$
\citep[e.g.,][]{Hamann18}.  There are several mechanisms for the
accelerated outflow winds, including radiative pressure from the
atomic lines and continuum \citep{Murray95, Proga00, Proga04,
  Proga12}, magneto centrifugal-force \citep{Everett05}, and thermal
pressure \citep[e.g.,][]{Chelouche05}.  The outflow winds influence
the evolutions of quasar host galaxies as well as the IGM surrounding
them through (1) ejecting angular momentum from the quasar accretion
disk to facilitate accretion onto the super-massive black hole (SMBH;
\citealt{Murray95}), (2) transporting metals and energy from the
central engine to the host galaxies as well as the IGM and
contributing to the chemical evolution of them
\citep[e.g.,][]{Dunn10}, (3) feeding back energy and momentum to
inter-stellar medium (ISM) of the host galaxies to inhibit star formation
\citep[e.g.,][]{DiMatteo05}.

The intrinsic absorption lines in rest-frame UV spectra are usually
classified into three categories: broad absorption lines (BALs) with
FWHM $\geq$ 2,000~\kms, narrow absorption lines (NALs) with FWHM
$\leq$ 500~\kms, and an intermediate subclass (mini-BALs).  BALs and
mini-BALs are easily identified as intrinsic absorption lines because
their large line width cannot be reproduced by thermal broadening or
turbulence of intervening absorbers.  On the other hand, it is not the
case for NALs because a typical line width of intrinsic NALs are
almost same as that of cosmologically intervening NALs, so it is
difficult to classify them based on only visual inspection. There are
several signs with which we can identify them as intrinsic NALs,
including i) time variability of absorption profile and/or strength
\citep[e.g.,][]{Hacker13}, ii) partial coverage\footnote{Absorbers do
  not cover the background flux source completely along our
  sight-line.} \citep[e.g.,][]{Ganguly99} and iii)
line-locking\footnote{The red member of a doublet is aligned with the
  blue member of the following doublet.} \citep[e.g.,][]{Arav96}, and
so on \citep{Hamann97}. Based on partial coverage analysis,
\citet{Misawa07} found $\geq$~50\%\ of optically bright quasars have
at least one intrinsic NAL (see also \citealt{Simon12}), which is
larger than those for BALs ($\sim$10--20\%; e.g., \citealt{Trump06,
  Gibson09}) and mini-BALs ($\sim$5--10\%; e.g.,
\citealt{Rodriguez2009}).  These results suggests that roughly
$\sim$70\%\ of quasars have at least one BAL, mini-BAL, and/or
intrinsic NAL in their spectrum \citep{Hamann12}.  In particular, BAL
is divided into three subclasses further based on a variation of
detected absorption lines: HiBAL showing only relatively
high-ionization lines (e.g., \ion{O}{6}, \ion{N}{5}, \ion{Si}{4} and
\ion{C}{4}) as well as \lya, LoBAL showing low-ionization lines (e.g.,
\ion{Al}{2}, \ion{Al}{3}, and \ion{Mg}{2}) in addition to the
high-ionization lines above \citep{Voit93}, and an extreme class
(FeLoBAL) showing very low-ionized iron lines as well \citep[e.g.,
][]{Hazard87, Becker97}.

There are two possible interpretations for detectability of BALs ---
(1) the orientation scenario, which suggests that quasars can be
observed as BAL quasars only if our line of sight (LOS) to the
continuum source of the 
wind quasars passes through the outflow wind \citep[e.g.,][]{Elvis00,
  Ganguly01}, and (2) the evolution scenario, which suggests that BAL
quasars are in a particular (probably early) evolutionary stage and
obscured by dust \citep[e.g.,][]{Boroson92, Voit93, Farrah07}, or they
are in a state with high Eddington ratio and high black hole mass
$M_{\mathrm {BH}}$ \citep{Giustini19}.  While the evolution scenario
has been suggested for LoBAL (especially for FeLoBAL) quasars \citep
{Farrah07}, the orientation scenario is more likely for HiBAL quasars
because BAL quasars and non-BAL quasars have similar emission line
spectra except for BAL features \citep[e.g.,][]{Weymann91,
  Reichard03}.  Based on the orientation scenario, it is also
suggested that quasars with narrower intrinsic absorption lines (i.e.,
mini-BALs and intrinsic NALs) are viewed at a smaller inclination
angle relative to an axis of the accretion disk \citep[unified model,
  hereafter; e.g.,][]{Ganguly01, Hamann12} because in most cases,
quasars with intrinsic NALs do not show any excess of X-ray absorption
by warm absorbers that are located near the base of the outflow wind
\citep{Misawa08, Chartas09}.  The numerical simulations of
\cite{Proga00} also show that the fast and dense streams which
probably observed as BALs emerge at a larger inclination angle. Thus,
the unified model has often been used to understand the difference
between BALs, mini-BALs, and intrinsic NALs.  If the model is
applicable for all quasars, neither mini-BAL nor intrinsic NAL is not
expected to be detected in BAL quasar spectra. However,
\cite{Rodriguez2009} detected mini-BALs in the BAL quasar
spectra. Thus, an inclination angle toward which BAL or mini-BAL is
detected has some overlap between them.  For intrinsic NALs, it has
not been systematically tested if they are present in a same
inclination angle of BAL/mini-BAL quasar.  \citet{Ganguly01} reported
there is an enhanced probability of detecting associated NALs in BAL
quasar spectra, but those NALs are not classified into intrinsic and
intervening ones.

In this study, we will confirm if {\it intrinsic} NALs are present in
BAL/mini-BAL quasar spectra by separating intrinsic NALs from
intervening ones with high-resolution spectra, to locate the origin of
intrinsic NALs and test the unified model.  This paper is structured
as follows.  In Section~2, we present sample selection and data
reduction.  A classification method of intrinsic NALs from intervening
NALs is described in Section~3.  The results and discussion are
presented in Sections~4 and 5, respectively.  In Section~6, we
summarize our results.  Throughout the paper, we use a cosmology with
$H_{0}=75~${\kms} ${\mathrm{Mpc}}^{-1}$, ${\Omega}_{\mathrm{m}}=0.3$,
and ${\Omega}_{\Lambda}=0.7$.

\section{Sample Selection $\&$ Data reduction}

For selecting sample quasars, we referred to the BAL quasar catalog in
\citet{Gibson09}. This catalog contains 5039 BAL quasars that are
identified based on modified Balnicity Index (BI$_0$) in any ions of
\ion{C}{4} {\lam}{\lam}1548,1551, \ion{Si}{4} {\lam}{\lam}1394,1403,
\ion{Al}{3} {\lam}{\lam}1855,1863 and \ion{Mg}{2}
    {\lam}{\lam}2796,2803, from the SDSS DR5 quasar catalog.  BI$_0$
    is essentially the same as traditional Balnicity Index (BI;
    \citealt{Weymann91}), but integrate from $0$\kms instead of the
    traditional $3,000~$\kms.  Among the 5039 BAL quasars in
    \citet{Gibson09}, 25 BAL/mini-BAL quasars with Dec $<
    +~15^{\circ}$ has been observed with VLT/UVES and already open to
    public in the archive, of which we use 11 BAL quasars as
    listed in Table~\ref{tab:SDSSsample} that satisfy all criteria
    below.  First, quality of spectra should be high enough after
    combining all spectra taken within 7~days (spectral resolution of
    $R$ $\geq$ 30,000 and median signal-to-noise ratio of S/N $\geq$ 6
    ).  Next, spectra should cover wavelength regions from
    \lya\ emission lines to one of \ion{C}{4}, \ion{Si}{4}, and/or
    \ion{N}{5} \footnote{We do not use \ion{Mg}{2} doublet because
      \citet{Jones10} show that low-ionized absorbers like \ion{Mg}{2}
      absorbers in molecular clouds could show partial coverage, even
      if they are intervening absorbers.} emission lines.
 If quasars have been observed in
multiple epochs, we choose the one that have highest S/N ratio and
largest wavelength coverage. All spectra in our sample are summarized
in Table~\ref{tab:alldata}.

\begin{table*}
\centering
\caption{Sample Quasars.}
\begin{center}
\begin{threeparttable}[h]
\centering
\begin{tabular}{ccccccccr} \hline
QSO name  & RA (hh:mm:ss) & DEC (dd:mm:ss) & $z_{\rm em}$ & $m_{\rm g}$ (mag)\tnote{a} & $m_{\rm r}$ (mag)\tnote{a} &  ${\mathcal R}$\tnote{b} & L/Q\tnote{c} & $\mathrm{BI_{0}}$ (\kms) \\ \hline
\ SDSS J022844.09+000217.0 & 02:28:44.1 & +00:02:16.8 & 2.72 & 18.45 & 18.22 & <3.08 & Q &1962.9\\
\ SDSS J024221.87+004912.6 & 02:42:21.9 & +00:49:12.0 & 2.06 & 18.50 & 18.20 & <3.66 & Q & 896.2\\
\ SDSS J110855.47+120953.3 & 11:08:55.5 & +12:09:53.3 & 3.67 & 20.14 & 18.60 & <5.03 & Q  & 213.0\tnote{d} \\
\ SDSS J115122.14+020426.3 & 11:51:22.1 & +02:04:26.4 & 2.40 & 19.06 & 18.60 & <6.70 & Q & 156.0 \\
\ SDSS J115944.82+011206.9 & 11:59:44.8 & +01:12:07.2 & 2.00 & 17.58 & 17.24 & 450    & L & 937.9\\
\ SDSS J120550.19+020131.5 & 12:05:50.2 & +02:01:30.0 & 2.13 &17.46 & 17.08 & <1.50 & Q & 403.9\tnote{d}\\
\ SDSS J120917.93+113830.3 & 12:09:17.9 & +11:38:27.6 & 3.11 & 17.64 & 17.47 & <1.76 & Q & 323.5\\
\ SDSS J121549.80$-$003432.1 & 12:15:49.8 & -00:34:30.0 & 2.71 & 17.50 & 17.35 & <1.76 & Q & 4807.7\\
\ SDSS J122848.21$-$010414.5 & 12:28:48.2 & -01:04:12.0 & 2.66 & 18.30 & 18.17 & 106  & L & 17.1\\
\ SDSS J133701.39$-$024630.3 & 13:37:01.4 & -02:46:30.0 & 3.06 & 19.08 & 18.70 &  259 & L & 2.3\\
\ SDSS J143907.51$-$010616.7 & 14:39:07.5 & -01:06:14.4 & 1.82 & 19.30 & 18.98 & <8.2 & Q &81.5\\ \hline 
\end{tabular}
\begin{tablenotes}
\item[a] $g$ or $r$-band magnitude referred from SDSS DR5.
\item[b]  Radio-loudness parameter, as defined in Section~2. We
  refer radio observation from FIRST survey.  If radio source is
  undetected, we use the detection limit of the FIRST survey as an
  upper limit of the radio flux.
\item[c]  Classification of radio-loudness: radio-loud (L) or
  radio-quiet (Q), as defined in Section~2.
\item[d] {No clear BAL features are detected, although this is
    classified into BAL quasars in \citet{Gibson09}.}
\end{tablenotes}
\end{threeparttable}
\end{center}
\label{tab:SDSSsample}
\end{table*}

\begin{table*}
\caption{Sample Quasar spectra taken from the ESO Archive.}

\begin{center}
\centering
 \begin{threeparttable}
\begin{tabular}{ccccrrr} \\ \hline \hline
QSO name & program ID & Wavelength range (\AA) & Date  & S/N\tnote{$\dag$} & Resolution   \\ \hline\hline 
\ SDSS J022844.09+000217.0 & 081.A-0479(A)   & 4728-6838$\mathrm{\AA}$ & 2008/08/29 & 11.7 & 42310\\ \hline
\ SDSS J024221.87+004912.6 & 075.B-0190(A)    & 3283-4564$\mathrm{\AA}$ & 2005/09/05 & 7.1 & 40970\\
\                                     &              &                                           &  2005/09/06         & 8.2 & 40970\\ 
\                                     &              &                                           &                              & 8.2 & 40970\\ 
\                                     &            & 3732-5000$\mathrm{\AA}$ & 2005/09/05 & 13.1 & 40970\\ 
\                                     &              &                                           &                             & 13.5 & 40970\\
\                                     &              &                                           &                              & 12.2 & 40970\\
\                                     &             & 5709-9464$\mathrm{\AA}$ & 2005/09/05 & 7.0 & 42310\\ 
\                                     &              &                                           &                             & 7.2 & 42310\\ 
\                                     &              &                                           & 2005/09/06 & 6.4 & 42310\\ \hline 
\ SDSS J110855.47+120953.3 & 083.A-0042(A) &4583-6687$\mathrm{\AA}$ & 2009/04/16 & 7.6 & 34540 \\
\                                     &            &                                             &                            & 6.7 & 34540 \\
\                                     &            &                                             & 2009/04/19         & 7.2 & 34540 \\
\                                     &            &                                             &                            & 7.1 & 34540 \\
\                                     &            &                                             & 2009/04/20         & 7.4 & 34540 \\ \hline
\ SDSS J115122.14+020426.3 & 092.B-0574(A) & 3732-5000$\mathrm{\AA}$ & 2014/03/05 & 8.3 & 36840 \\
\                                     &            &                                             &                            & 8.1 & 36840 \\ 
\                                     &            &                                             & 2014/03/06 & 7.6 & 36840 \\
\                                     &            & 4583-6687$\mathrm{\AA}$  & 2014/02/24 & 11.1 & 34549 \\
\                                     &            &                                             & 2014/02/26 & 11.2 & 34540 \\
\                                     &            &                                             &  2014/02/27 & 11.8 & 34540 \\ \hline
\ SDSS J115944.82+011206.9 & 079.B-0469(A) & 3703-5054$\mathrm{\AA}$  & 2007/06/06 & 15.1 & 40970 \\
\                                     &            &                                             & 2007/06/07         & 12.0 & 40970 \\
\                                     &         & 5603-9612$\mathrm{\AA}$   & 2007/06/06 &9.3 & 42310 \\
\                                     &            &                                             & 2007/06/07       & 7.4 & 42310 \\ \hline
\ SDSS J120550.19+020131.5 & 273.A-5020(A) & 3282-4564$\mathrm{\AA}$ & 2004/05/17 & 13.1 & 40970\\
\                                     &         & 4728-6837$\mathrm{\AA}$ &                        & 14.2 & 42310\\ \hline
\ SDSS J120917.93+113830.3 & 080.A-0482(A) & 4959-7071$\mathrm{\AA}$ & 2006/02/01 & 7.9 & 42310\\ \hline
\ SDSS J121549.80$-$003432.1 & 185.A-0745(D) & 3282-4563$\mathrm{\AA}$ & 2011/03/28 & 6.4 & 49620\\
\                                     &           &                                           &                                 & 7.6 & 49620\\
\                                     &           &                                            & 2011/03/29              & 7.8 & 49620\\
\                                     &           &                                           &                                 & 7.4 & 49620\\
\                                     &          & 4583-6687$\mathrm{\AA}$ & 2011/03/28 & 13.3 & 51690\\
\                                     &           &                                           &                             & 15.1 & 51690\\
\                                     &           &                                           & 2011/03/29 & 14.5 & 51690\\
\                                     &           &                                           &                            &  15.0 & 51690\\
\                                     &          & 4727-6835$\mathrm{\AA}$ & 2011/03/30 & 7.6 & 51690\\
\                                     &            &                                            &                            & 12.0 & 51690\\
\                                     &            &                                            & 2011/03/31          & 13.8 & 51690\\
\                                     &            &                                            &                            & 11.0 & 51690\\
\                                     &            &                                            & 2011/04/01           & 11.4 & 51690\\
\                                     &            &                                            &                             & 12.3 & 51690\\ \hline
\ SDSS J122848.21$-$01041.5 & 081.A--0334(A) & 3259-4519$\mathrm{\AA}$ & 2008/04/07 & 6.2 & 40970\\
\                                     &         & 4727-6835$\mathrm{\AA}$ &                       & 12.1 & 42310\\ \hline
\ SDSS J133701.39$-$024630.3 & 091.A--0018(A) & 4583-6687$\mathrm{\AA}$ & 2014/02/09 & 11.7 & 37820\\
\                                        &           &                                            & 2014/02/10 & 11.9 & 37820\\
\                                        &           &                                            &                     & 11.8 & 37820\\
\                                        &           &                                            & 2014/02/11 & 12.0 & 37820\\ \hline
\ SDSS J143907.51$-$010616.7 & 081.B-0285(A) & 3732-5000$\mathrm{\AA}$ & 2008/05/01 & 5.6 & 40970\\
\                                        &           &                                              & 2008/05/02 & 7.4 & 40970\\ \hline
\end{tabular}
\label{tab:alldata}
\begin{tablenotes}\footnotesize
 \item[\dag]  Median signal to noise ratio per exposure before combining other spectra.
 \end{tablenotes}
 \end{threeparttable}
 \end{center}
\end{table*}

We retrieved raw data from the ESO
archive\footnote{http://archive.eso.org} and reduced them ourselves
using the ESO Reflex workflow.  The ESO archive provides both raw and
reduced data so that we can carefully check a quality of spectra by
comparing spectra reduced by the ESO archive and those by ourselves.
After that, we applied a helio-centric velocity correction and an
air-to-vacuum wavelength correction.  If there are multiple spectra
taken with same observing configuration within 7~days in the observed
frame, we combine all of them as a single epoch to improve S/N ratio.
We adopt 7~days as the maximum time interval for a single epoch
because the shortest time scale of BAL variability in quasar
rest-frame is a week to a month \citep{Capellupo11}.  Although
\ion{C}{4} BALs sometimes exhibit significant equivalent width
variability in a time scales of $<$ 10~days, the fraction of epoch
pair that exhibit significant BAL variability is very small
($\sim3.7$\%, \citealt{Hemler19}).  Finally, we fitted the continuum
and broad emission lines with a cubic spline function to make
normalized spectra, except for heavily absorbed regions near emission
lines (i.e., P-Cygni like regions).
 Based on our visual inspection, we found two quasars
  (SDSS~J110855.47+120953.3 and SDSS~J120550.19+020131.5) did not show
  clear BAL features, although these had been classified into BAL
  quasars. We will discuss these later in Section 4.

 We also summarize radio-loudness of our sample quasars in
  Table~\ref{tab:SDSSsample} .  The Radio-loudness is defined as a
  ratio of the flux densities at ${\rm 5~GHz}$ and 4400 \AA, i.e.,
  $\mathcal{R}=f~( {\rm 5~GHz})~ /~ f~(4400)$. We derived  $f(4400)$
  from the SDSS $g$-band magnitude $m_{\rm g}$ assuming optical
  spectral index $\alpha_{\rm o}= 0.44$ for all quasars, except for
  three (SDSS~J110855.47+120953.3, SDSS~J120917.93+113830.3
  and SDSS~J133701.39-024630.3 at $z_{\rm em}>3$), for which we used
  $r$-band magnitude $m_{\rm r}$ to avoid \lya~forest. As for the
  radio flux, we derived $f~({\rm 5~GHz})$ from FIRST ${\rm1.4 ~GHz}$
  survey assuming a radio spectral index of $\alpha_{\rm
    r}=0.7$. According to the criterion of radio-loudness (i.e.,
  ${\mathcal R}>10$ as introduced by \citealt{Kellermann89}), three
  out of the 11 BAL quasars are classified as radio-loud quasars.
 The fraction of radio-loud quasars among our sample is about 27\%, which 
 is consistent with the fraction among ordinary quasars \citep[$\sim$10-20\%; ][]
 {Kellermann89, Urry95, Ivezic02} and also weakly consistent with that of BAL 
 quasars (e.g., \citealt{Tolea02}), although it seems to depend on both luminosity 
 and redshift (\citealt{Jiang07}; \citealt{Banados15}).

\section{Analysis}
\label{sec:analysis}
In this section, we first search for NALs in the 11 BAL quasar spectra
(Sec. 3.1), and classify them into intrinsic and intervening NALs
(Sec. 3.2 \& 3.3) using partial coverage analysis.

\subsection{Detection of NALs}
We search for all of \ion{C}{4} {\lam}{\lam}1548,1551, \ion{Si}{4} {\lam}{\lam}1394,1403, 
and \ion{N}{5} {\lam}{\lam}1239,1243 NALs that
are detected with a confidence level of $> 3 \sigma$; i.e.,
$[1-R_{\rm abs}]/\sigma(R_{\rm abs}) > 3$, where
$R_{\rm abs}$ and $\sigma(R_{\rm abs})$ are residual flux and its
uncertainty at the bottom of absorption line in the normalized
spectrum.  We also combine all NALs within 200~\kms\ each other into a
single absorption {\it system} as done in \citet{Misawa07} because
they are probably not physically independent.  Finally, we prepare a
homogeneous sample for statistical analyses below by removing NALs
that are i) detected near P-Cygni profile (because continuum fitting
is less reliable), ii) detected at low S/N region smaller than 6
pix$^{-1}$ (because line depth is less reliable), and/or iii) having
equivalent width ($ \rm EW$) smaller than $4\sigma(\rm EW)$, where
$\sigma(\rm EW)$ is $1\sigma$ uncertainty of equivalent width.  We
obtain 40 \ion{C}{4} and 12 \ion{Si}{4} NALs but no \ion{N}{5} NAL in
45 NAL systems detected in the 11 BAL/mini-BAL quasar spectra, as
listed in Table~\ref{tab:resland}.

\subsection{Partial coverage analysis}
We use a partial coverage analysis to identify intrinsic NALs rather
than time-variability or line-locking analysis because i) it requires
only a single epoch spectrum while time variability analysis
requires multiple epoch spectra and ii) NALs with partial coverage are
frequently detected in $\sim50$\%\ of quasars while a line locking is
very rare phenomena. Optical depth ratio of doublet lines by
Lithium-like ions such as \ion{C}{4}, \ion{Si}{4} or \ion{N}{5} should
be 2:1 based on atomic physics.  However, if absorbers do not cover
the background flux source entirely along our LOS (i.e.,
partial coverage), the optical depth ratio would deviate from 2:1
\citep{Hamann97, Barlow97}.  We can evaluate a fraction of flux that
is covered by the absorber (i.e., Covering factor, $C_{\rm f}$) by
measuring residual flux of doublet lines by
\noindent 
\begin{equation}
\label{eq1}
 C_{\rm f}(v)=\frac{(1-R_{\rm r}(v))^{2}}{R_{\rm b}(v)-2R_{\rm r}(v)+1},
\end{equation}
where $R_{b}$ and $R_{r}$ are residual flux of blue and red members of
doublet. If $C_{\rm f}$ is smaller than unity, the corresponding
absorber is probably associated to the outflow winds, because the size
of another possible candidate (i.e., foreground galaxies and
intergalactic medium) is larger than the flux source by several orders
of magnitude.  We measure $C_{\rm f}$ by fitting absorption profile of
NALs with Voigt profiles using the software package \textsc{minfit}
(\citealt{Churchill01}, hereafter, fitting method). The
\textsc{minfit} is a line fitting code, with which we can performs a
chi-square fitting using absorption redshift ($z_{\rm abs}$), column
density ($\log N$), Doppler parameter ($b$), and Covering factor
($C_{\rm f}$) as free parameters.  In the fitting process, we
convolved a model profile with the instrumental line spread function.
During the fitting trials, the code sometimes gives unphysical values
of $C_{\rm f}$ such as $C_{\rm f} > 1$ or $C_{\rm f} < 0$, which tends
to happen if absorption depth is very shallow and/or a real $C_{\rm
  f}$ value is close to unity. In such cases, we repeat a Voigt
profile fitting assuming full coverage (i.e., $C_{\rm f} = 1$) with no
error bar for those components with unphysical $C_{\rm f}$ values in
the first fitting trials as done in \citet{Misawa07}.  In addition to
the fitting method, we also measure $C_{\rm f}$ value for each {\it
  pixel} (hereafter, pixel-by-pixel method) to confirm the result of
fitting method.  Covering factor measured for each pixel is less
  reliable compared to that by the fitting method since the former is
  easily affected by photon noise as well as an instrumental line
  spread function in the wing parts of absorption trough
  \citep{Ganguly99}. Therefore, we mainly use the results from the
  fitting method, while we use the pixel-by-pixel method in a
  supplementary way only if the fitting method does not give
  conclusive results.

\subsection{Classification of NAL systems}

We classified 52 NALs in 45 systems into three classes: class~A
(reliable intrinsic NALs), class~B (possible intrinsic NALs), and
class~C (intervening or unclassified NALs) based on results of the
fitting and the pixel-by-pixel methods, which is the same
classification as introduced by \cite{Misawa07}.
We classify all NALs into class~A if they show a partial coverage with
3$\sigma$ confidence level (i.e., $ C_{\rm f} + 3 \sigma(C_{\rm f}) <
1$) based on the fitting method.
Among NALs that do not satisfy the criteria of class~A, we search for
class~B NALs (i.e., possible intrinsic but less likely compared to
class~A) that satisfy the following criterion: they show partial
coverage with 1$\sigma$ confidence level (i.e., $ C_{\rm f} +
\sigma(C_{\rm f})< 1$) based on the fitting method as
well as the pixel-by-pixel method at the center.
The other NALs are classified into class~C that contains subclasses
below: NALs consistent with full coverage (i.e., $ C_{\rm f} +
\sigma(C_{\rm f}) >1$) for all components based on the fitting method
as well as the pixel-by-pixel method (class~C1), unclassified NALs
because they are weak and blending with stronger components
(class~C2), and NALs for which we cannot apply reliable model fit
because of continuum fitting and/or data defect such as blending with
physically unrelated lines and noise spikes (class C3).

Among the five classes including subclasses, classes~A, B, and C1 are
those that are classified based on the partial coverage analysis,
while classes~C2 and C3 are unclassified NALs because of severe line
blending and/or low quality of the spectrum. We also apply the same
classification for NAL {\it systems} as follows. If NAL systems
contain at least one component that is classified into class~A,
we classify the system as a whole into class~A regardless of the existence 
of class~B and C NALs. In the same way, if NAL systems have at least 
one class B NAL but no class~A NALs, we classify them into class B as a 
whole even if the other components are consistent with full coverage.

\section{Results}

Using partial coverage analysis, we classify 45 NAL systems into 1
class~A, 2 class~B, and 42 class~C systems, of which 29 are class~C1,
3 are class~C2, and 10 are class~C3.  Even if we remove two
  quasars without clear BAL features (as noted in Section 2), we still
  detect 1 class A and 2 class B NALs but smaller number (33) of
  class C NALs in 36 NAL systems. Thus, in BAL quasar
spectra we detect three candidates for {\it intrinsic} NAL
systems. All of them are identified with \ion{C}{4} NALs. In spite of
rather small sample size, our result suggests that intrinsic NAL
absorbers locate along our LOS to $\sim33^{+33}_{-18}$\%\ (3/9) of BAL
quasars.  Here, we would like to emphasize that the fraction is a lower limit
because intrinsic NALs do not always exhibit partial coverage.

  Table~\ref{tab:resland} summarizes parameters of 36 NAL systems
  in 9 BAL/mini-BAL quasars with clear BAL features as well as 9 NAL
  systems in two quasars without clear BAL features, including quasar
  name, emission redshift, ion name, wavelength in the observed frame,
  equivalent width in the observed frame, absorption redshift, offset
  velocity, column density, Doppler $b$ parameter, covering fraction,
  NAL class, and other transitions of absorption lines detected in a
  same NAL system. Since emission redshift is sometimes blueshifted
  from the systemic redshift, offset velocities in the table could be
  underestimated by up to a few hundreds of \kms. On the other hand,
  our main fit parameter (i.e., covering factors) is not affected by
  the effect and still reliable.


\begin{landscape}
\begin{table}
 \caption{Properties of NALs, including quasar name in SDSS DR5
   (QSO~name), emission redshift (${z_{\mathrm{em}}}$), transition of
   absorption line (ION), observed-frame wavelength corresponding to a
   blue member of doublet ($\mathrm{\lambda_{obs}}$), observed-frame
   equivalent width of a blue member of doublet ($\mathrm{EW_{obs}}$),
   absorption redshift (${z_{\mathrm{abs}}}$), offset velocity from
   background source ($v_{\mathrm {off}}$: blueshift is positive),
   column density (logN), Doppler b
   parameter ($b$), covering
   fraction $C_{\mathrm{f}}$, NAL class (Class), and a list of other
   transitions of absorption lines that are detected at the same
   redshift (Other ions). Transitions in parenthesis are lines detected 
   in the \lya -forest (i.e., less reliable). For class C3 systems, 
   we present only $ {EW_{\mathrm{obs}}}$, ${z_{\mathrm{abs}}}$ and ${v_{\mathrm{off}}}$  
    since we cannot place any meaningful constraints on fit parameters.}
 \label{tab:resland}
 \begin{threeparttable}
\begin{tabular}[h]{lccccccccccc}
  \hline
QSO name                             & ${z_{\mathrm{em}}}$  &  ION   & $\mathrm{\lambda_{obs}}$  & $ {EW_{\mathrm{obs}}}$ & ${z_{\mathrm{abs}}}$  & ${v_{\mathrm{off}}}$   &  $\log N$   &  $b$ & $C_{\mathrm {f}}$\tnote{\dag} & Class  & Other ions\\  
                                               &                                  &            &   ({\AA})                                &        ({\AA})                  &                                   & (\kms)                      &  (\cmm)   &        (\kms)                     &                       &             &               \\           \hline\hline
\ SDSS J022844.09+000217.0 & 2.72  & \ion{C}{4}  & 5192.1  & 0.730  & 2.353     & 30994 &                               &                        &                   &   C1  & (\lya,  \ion{Si}{3} {\lam} 1207,  \ion{Si}{2} {\lam} 1193)        \\ 
\                                                &           &                  &               &             & 2.3536 & 30998 & $13.75\pm0.02$    & $37.1\pm2.2$    &  1.00                               &                &                     \\
\                                                &           &                  &               &             & 2.3535 & 31008 & $13.58\pm0.18$    & $11.7\pm1.6$   & $0.79^{+0.54}_{-0.31}$           &           &      \\ \cline{3-12}
\                                                &           &  \ion{Si}{4}& 4892.5 & 0.934  & 2.446      & 22929 &                              &                           &                                                 &     C3         &    (\lya,  \ion{Si}{3} {\lam} 1207,  \ion{Si}{2} {\lam} 1193)     \\        
\                                                &           & \ion{C}{4}  & 5335.0   & 2.366   & 2.446   & 22912  &                                &                        &                                                 &       C1        &        \\
\hline

 \end{tabular} 
 \end{threeparttable}
\end{table}
\end{landscape}

\begin{landscape}
\begin{table}
 \contcaption{A table continued from the previous one.}
 \label{tab:continued}
  \begin{threeparttable}
   \begin{tabular}[h]{lccccccccccc}
  \hline
QSO name                             & ${z_{\mathrm{em}}}$  &  ION   & $\mathrm{\lambda_{obs}}$  & ${EW_{\mathrm{obs}}}$ & ${z_{\mathrm{abs}}}$  & ${v_{\mathrm{off}}}$   & $\log N$   & $b$ & $C_{\mathrm {f}}$\tnote{\dag} & Class  & Other ions\\  
                                               &                                  &            &   ({\AA})                                &        ({\AA})                  &                                   & (\kms)                      &             &        (\kms)                     &                       &             &               \\           \hline\hline
\ SDSS J110855.47+120953.3 & 3.67 & \ion{C}{4}     & 6218.8   & 0.344  & 3.017     & 44863 &                               &                         &                   &   C3  &      \\  \cline{3-12}
\                                                &           & \ion{C}{4}     & 6251.7   & 1.665  & 3.038     & 43314 &                               &                         &                   &   C3  &   (\ion{Si}{3} {\lam} 1207, \lya)   \\  \cline{3-12}
\                                                &           & \ion{C}{4}     & 6292.0   & 2.282  & 3.063     & 41495 &                               &                         &                   &   C3  &   (\lya)  \ion{C}{2} {\lam} 1335    \\  \cline{3-12}
\                                                &           & \ion{Si}{4}     & 6335.0   & 1.275  & 3.545     & 8117 &                               &                         &                   &   C1  &  (\ion{Si}{3} {\lam} 1207, \lya)   \\  
\                                                &           &                     &               &              & 3.5449  & 8141 & $13.11\pm0.02$  & $20.3\pm1.1$     & 1.00                                               &                        &  \ion{Si}{2}  {\lam} 1260, \ion{Si}{2} {\lam} 1304, \ion{C}{2} {\lam} 1335     \\
\hline

 \end{tabular} 
  \end{threeparttable}
\end{table}
\end{landscape}

\begin{landscape}
\begin{table}
 \contcaption{A table continued from the previous one.}
 \label{tab:continued}
  \begin{threeparttable}
   \begin{tabular}[h]{lccccccccccc}
  \hline
QSO name                             & ${z_{\mathrm{em}}}$  &  ION   & $\mathrm{\lambda_{obs}}$  & ${EW_{\mathrm{obs}}}$ & ${z_{\mathrm{abs}}}$  & ${v_{\mathrm{off}}}$   &  $\log N$   & $b$ & $C_{\mathrm {f}}$\tnote{\dag} & Class  & Other ions\\  
                                               &                                  &            &   ({\AA})                                &        ({\AA})                  &                                   & (\kms)                      &             &        (\kms)                     &                       &             &               \\           \hline\hline

\ SDSS J115944.82+011206.9 & 2.00  & \ion{C}{4}  & 4088.6  & 0.370      & 1.641   & 38014 &                               &                            &                                                   &      B         &           \\
\                                                &          &                   &              &                & 1.6409 & 38014 & $13.94\pm0.03$     & $12.3\pm0.3$     & $0.87^{+0.06}_{-0.06}$            &                   &          \\ \cline{3-12}
\                                                &          & \ion{Si}{4}  & 3790.4   & 0.313     & 1.720   & 29316  &                               &                             &                                                  &   C3           &         \\
\                                                &          & \ion{C}{4}   & 4211.2    & 1.475    & 1.720    & 29265 &                               &                              &                                                 &    C1          &          \\
\                                                &          &                   &                 &             & 1.7193  & 29352 & $14.02\pm0.03$    & $21.9\pm0.8$       & $0.91^{+0.05}_{-0.04}$           &                   &        \\
\hline
 \end{tabular} 
  \end{threeparttable}
\end{table}
\end{landscape}

\begin{landscape}
\begin{table}
 \contcaption{A table continued from the previous one.}
 \label{tab:continued}
  \begin{threeparttable}
   \begin{tabular}[h]{lccccccccccc}
  \hline
QSO name                             & ${z_{\mathrm{em}}}$  &  ION   & $\mathrm{\lambda_{obs}}$  & ${EW_{\mathrm{obs}}}$ & ${z_{\mathrm{abs}}}$  & ${v_{\mathrm{off}}}$   &  $\log N$   & $b$ & $C_{\mathrm {f}}$\tnote{\dag} & Class  & Other ions\\  
                                               &                                  &            &   ({\AA})                                &        ({\AA})                  &                                   & (\kms)                      &             &        (\kms)                     &                       &             &               \\           \hline\hline
\ SDSS J120917.93+113830.3 & 3.11  & \ion{C}{4}   & 5146.0    & 0.376    & 2.324   & 62746  &                                 &                             &                                                  &     C3            &    ( \ion{Si}{2} {\lam} 1193, {\lya} , \ion{C}{2} {\lam} 1335 )   \\   \cline{3-12}
\                                                &          & \ion{C}{4}   & 5495.6     & 2.410   & 2.550  & 43658 &                                   &                             &                                                   &   C1               &         (\lya)      \\
\                                                &          &                   &                 &             & 2.5490 & 43718 & $14.64\pm1.85$       & $6.0\pm5.0$        & $0.52^{+0.17}_{-0.17}$             &                      &               \\        
\                                                &          &                   &                 &             & 2.5494 & 43682 & $14.02\pm0.30$       & $9.4\pm3.0$        & $0.95^{+0.25}_{-0.25}$             &                      &             \\
\                                                &          &                   &                 &             & 2.5495 & 43673 & $14.37\pm0.13$       & $52.0\pm9.2$      & $0.92^{+0.18}_{-0.16}$              &                     &             \\
\hline

 \end{tabular}
 \end{threeparttable}
\end{table}
\end{landscape}

 \begin{landscape}
\begin{table}
 \contcaption{A table continued from the previous one.}
 \label{tab:continued}
  \begin{threeparttable}
   \begin{tabular}[h]{lccccccccccc}
  \hline
QSO name                             & ${z_{\mathrm{em}}}$  &  ION   & $\mathrm{\lambda_{obs}}$  & ${EW_{\mathrm{obs}}}$ & ${z_{\mathrm{abs}}}$  & ${v_{\mathrm{off}}}$   &  $\log N$  & $b$ & $C_{\mathrm {f}}$\tnote{\dag} & Class  & Other ions\\  
                                               &                                  &            &   ({\AA})                                &        ({\AA})                  &                                   & (\kms)                      &             &        (\kms)                     &                       &             &               \\           \hline\hline
\ SDSS J121549.80-003432.1 & 2.71   & \ion{Si}{4}   & 4991.5    & 0.607    & 2.581   & 10583 &                                 &                             &                                                  &       C3           & (\lya )   \\   
\                                                &          &                    &                &              &             &            &                                 &                               &                                                   &                        & \ion{C}{4} {\lam}{\lam} 1548,1551 \\ \cline{3-12}
\                                                &          & \ion{C}{4}    & 5685.4    & 0.464    & 2.672   & 3068   &                                 &                               &                                                  &    C1            &  (\ion{H}{1} {\lam} 1026, \ion{O}{6} {\lam}{\lam} 1032.1038 \\ 
\                                                &          &                    &                &              & 2.6716 & 3118   & $13.11\pm0.08$      & $5.7\pm0.7$           & 1.00                                        &                       &  \ion{Fe}{2} {\lam} 1062, \ion{Fe}{2} {\lam} 1125, \lya)\\
\hline

  \end{tabular}
   \end{threeparttable}
\end{table}
\end{landscape}

 \begin{landscape}
\begin{table}
 \contcaption{A table continued from the previous one.}
 \label{tab:continued}
  \begin{threeparttable}
   \begin{tabular}[h]{lccccccccccc}
  \hline
QSO name                             & ${z_{\mathrm{em}}}$  &  ION   & $\mathrm{\lambda_{obs}}$  & ${EW_{\mathrm{obs}}}$ & ${z_{\mathrm{abs}}}$  & ${v_{\mathrm{off}}}$   &  $\log N$   & $b$ & $C_{\mathrm {f}}$\tnote{\dag} & Class  & Other ions\\  
                                               &                                  &            &   ({\AA})                                &        ({\AA})                  &                                   & (\kms)                      &             &        (\kms)                     &                       &             &               \\           \hline\hline
\ SDSS J133701.39-024630.3 & 3.06  & \ion{C}{4}    & 5064.6 & 2.739    & 2.271   & 64194 &                                 &                                 &                                                  &       C1           &    ( \ion{Si}{2} {\lam} 1190, \ion{Si}{3} {\lam} 1207, \lya    \\   
\                                                &          &                    &                &              & 2.2710 & 64220 & $17.42\pm0.31$      & $19.3\pm1.0$       & $0.99^{+0.06}_{-0.06}$            &                       &    \ion{N}{5} {\lam}{\lam} 1239,1243, { } \ion{Si}{2} {\lam} 1260) \\
\                                                &          &                    &                &              & 2.2718 & 64148 & $13.89\pm0.04$      & $20.2\pm1.0$       & $0.97^{+0.07}_{-0.06}$            &                   &  \ion{C}{2} {\lam} 1335, \ion{Al}{2} {\lam} 1671 \  \\
\                                                &          &                    &                &              & 2.2723 & 64103 & $13.55\pm0.07$      & $8.7\pm0.4$         & $0.95^{+0.08}_{-0.08}$            &                    &     \ion{Al}{3} {\lam}{\lam} 1855,1863  \\
\hline
 \end{tabular}
 \begin{tablenotes}\footnotesize
 \item[\dag]  If the $C_{\mathrm {f}}$ error is larger than 1.0, we replace it with 1.00.
 \end{tablenotes}
 \end{threeparttable}
\end{table}
\end{landscape}

\subsection{Intrinsic NALs in BAL quasars}
We discovered intrinsic \ion{C}{4} NALs at $z_{\rm abs} =1.945$ in
SDSS J024221.87+004912.6 (hereafter J0242+0049), at $z_{\rm abs} =
1.641$ in SDSS J115944.82+011206.9 (hereafter J1159+0112) and at
$z_{\rm abs} =2.961$ in SDSS J121549.80-003432.1 (hereafter
J1215-0034).  Among these, an identification of the NAL system in
J0242+0049 is highly reliable because it shows a partial coverage with
sufficiently high confidence level of $>7\sigma$. We describe detailed
properties of the three intrinsic NAL systems below.

\subsubsection{Class~A NAL at $z_{\rm abs}=1.945$ in J0242+0049}
In the spectrum of J0242+0049 ($z_{\rm em} = 2.062$, $m_{\rm g}
=18.67$), there exist both a mini-BAL with an offset velocity from the
quasar emission redshift ($v_{\rm off}$) of $\sim 3,000$~\kms and a
BAL with $v_{\rm off} \sim 18,000$~\kms.  \citet{Hall07} studied the
mini-BAL and BAL systems in detail, and discovered that \ion{Si}{4}
NALs corresponding to the \ion{C}{4} mini-BAL show line-locking
between themselves. They also found the BAL is accelerated toward our
LOS from June 2001 (SDSS data) to September 2005 (VLT/UVES data) by
$0.154\pm0.025$~cm~s$^{-2}$.

\begin{figure}
	\includegraphics[width=\columnwidth]{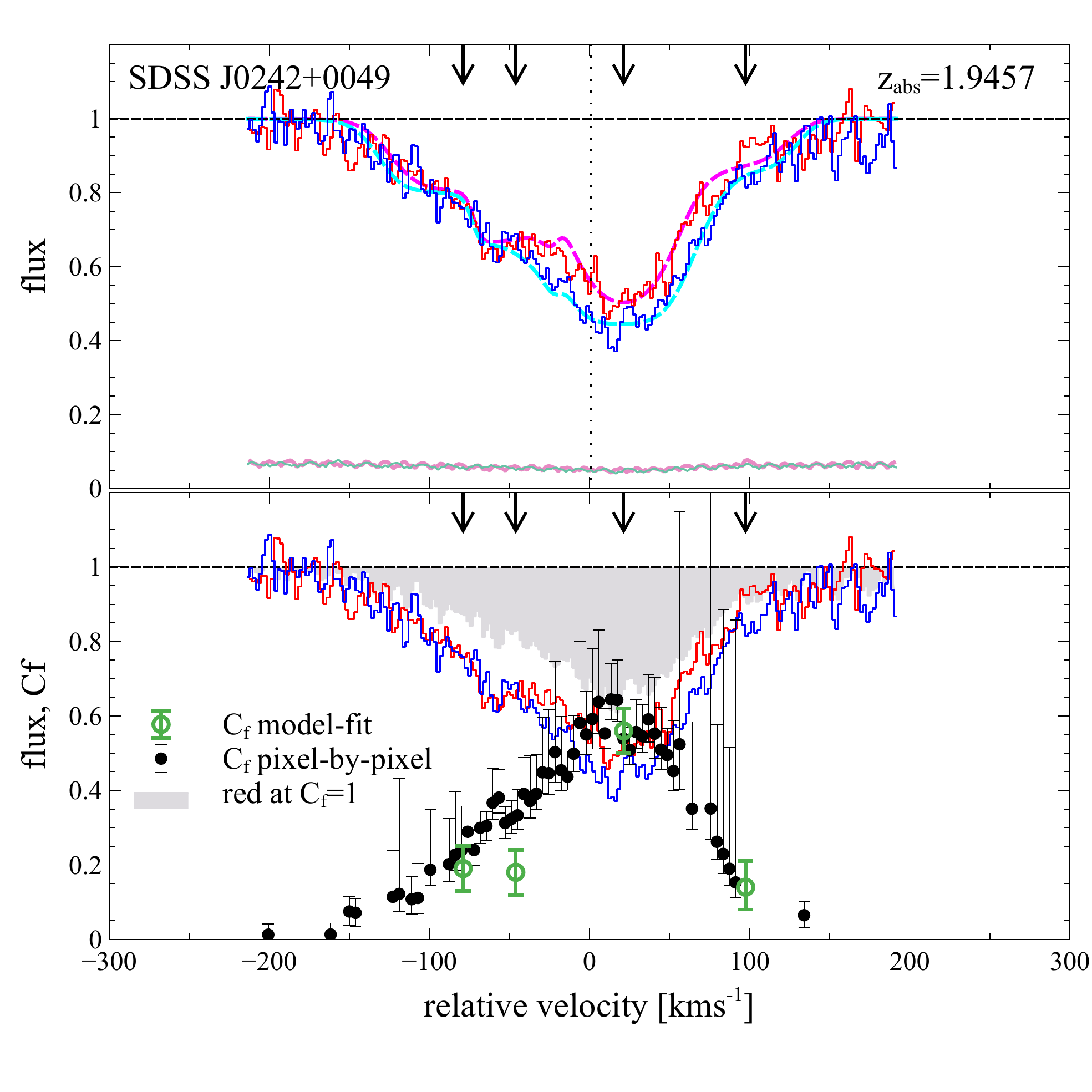}
	\caption{Best-fit model for \ion{C}{4} NAL at $z_{\rm
            abs}=1.945$ in BAL quasar J0242+0049. Downward arrows in
          both panels denote positions of absorption components. In
          the top panel, blue and red members of doublet are shown
          with blue and red histogram, while fitting results for them
          are shown with dashed cyan and magenta lines,
          respectively.  Flux weighted line center is shown with a
          vertical dotted line.  Bottom lines denote normalized
          $1\sigma$ errors.  In the bottom panel, covering factors
          evaluated for each component or each pixel are plotted with green
          open circles or black filled circles. Shaded gray profile
          denotes a profile of \ion{C}{4} $\mathrm{\lambda}$ 1551 that
          is expected from the observed profile of \ion{C}{4}
          $\mathrm{\lambda}$ 1548 assuming a full coverage.}
	\label{fig:J02421945}
\end{figure}

This quasar has been observed only once with VLT/UVES in September
2005. The observed spectrum covers $\lambda$ $\sim 3,300-5,000$~{\AA}
and $\sim 5,700-9,400$~{\AA} with a typical spectral resolution of $R$
$\sim 41,000$. The intrinsic \ion{C}{4} NAL at $z_{\rm em} =1.945$ is
detected at wavelength region with sufficiently high S/N ratio of
$\sim$17 pix$^{-1}$ for partial coverage analysis and
is not overlapped with other absorption/emission lines. Best-fit model 
to the \ion{C}{4} NAL in Fig.~\ref{fig:J02421945} gives partial coverage 
of $C_{\rm f}=0.56\pm0.06$ for the main component, which satisfies the
criterion of class~A (i.e., $C_{\rm f}+7\sigma(C_{\rm f})<1.0$).

\begin{figure*}
	\includegraphics[width=180mm]{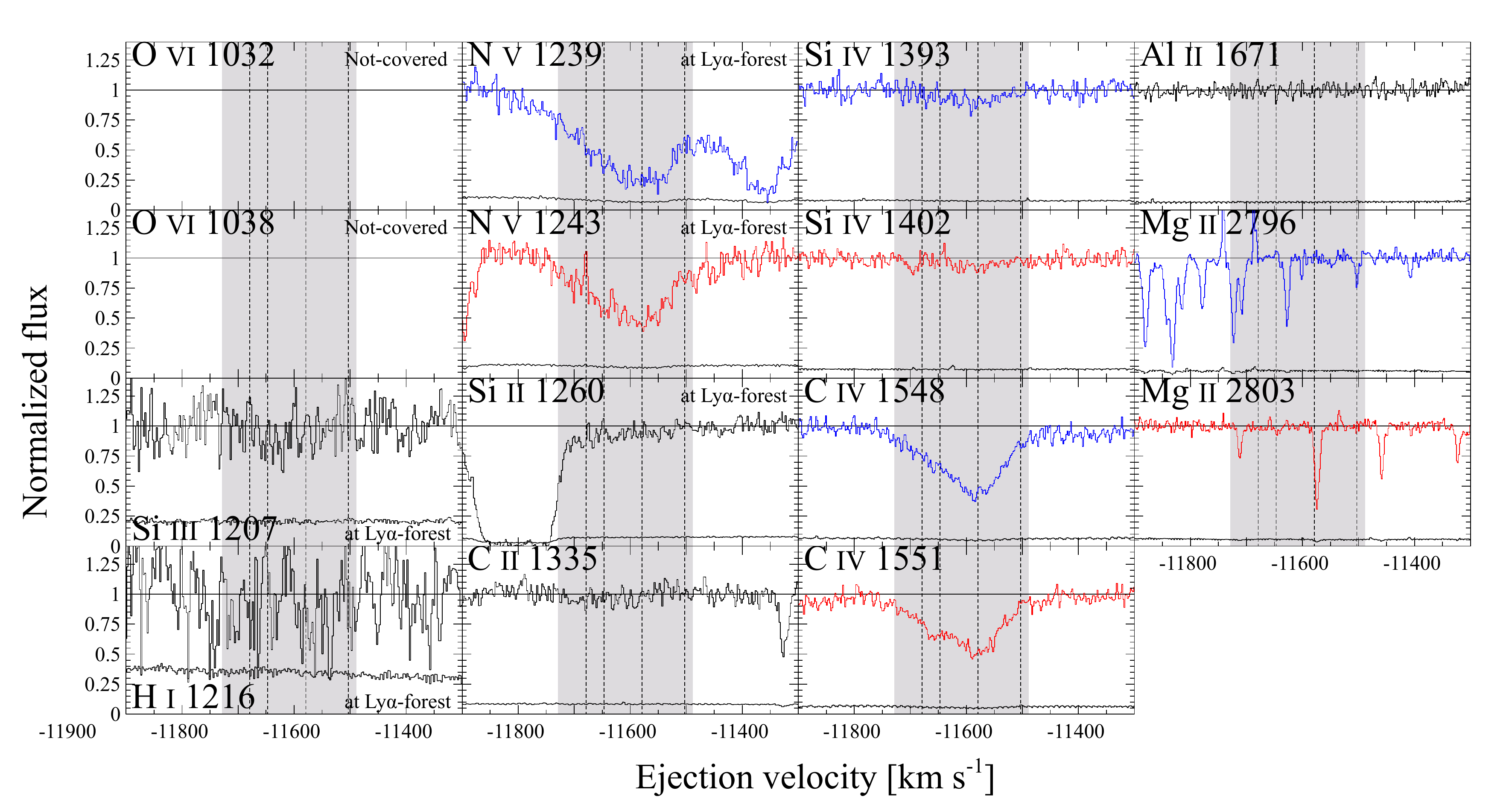}	
	\caption{Velocity plot of NAL system at $z_{\rm abs}=1.945$ in
          J0242+0049. Zero velocity corresponds to the quasar's
          emission redshift. Blue and red members of \ion{N}{5},
          \ion{Si}{4}, and \ion{C}{4} doublets are shown in blue and
          red histograms, respectively, while singlets are shown with
          black histograms.  Shaded gray regions show the line width
          of the broadest absorption line without line blending in the
          system (i.e., \ion{C}{4}), where a normalized flux remains
          below 0.9. Vertical dashed-lines corresponds to the
          positions of components of the \ion{C}{4} NAL.}
	\label{fig:J02421945sys2}
\end{figure*}

As shown in Fig.~\ref{fig:J02421945sys2}, high-ionization lines such
as \ion{N}{5}\footnote{We do not apply partial coverage analysis for 
the \ion{N}{5} NAL because it is heavily blending with \lya{} forest.} 
 with an ionization potential of IP = 98~eV and \ion{C}{4}
(IP = 64~eV) are obviously detected, while low-ionization lines such
as \ion{Al}{2} $\mathrm{\lambda}$1671 (IP = 19~eV), \ion{C}{2}
$\mathrm{\lambda}$1335 (IP = 24~eV), and \ion{Mg}{2}
$\mathrm{\lambda}$$\mathrm{\lambda}$2796,2803 (IP = 15~eV) are
absent. An extremely high-ionized \ion{O}{6}
$\mathrm{\lambda}$$\mathrm{\lambda}$1032,1038 line (IP = 138~eV) is
not covered by the observed spectrum.

Since both \ion{N}{5} and \ion{C}{4} NALs have smooth line profile
with large line width (the flux-weighted line width of $\sigma_{\rm v}
\geq 100~$\kms) that cannot be explained by thermal broadening only,
their absorbers should have an external turbulence, which could be
caused by a radiative pressure. Indeed, a large offset velocity of the
NAL system from the quasar ($v_{\rm off} \sim 11,600$~\kms) supports
that the corresponding absorber is an effectively accelerated outflow 
wind as frequently reported in the literature (\citealt{Ganguly07}, \citealt{Wild08},
\citealt{Misawa14}).

\subsubsection{Class~B NAL at $z_{\rm abs}=1.641$ in J1159+0112}
J1159+0112 ($z_{\rm em} =1.999$, $m_{\rm g} = 17.59$), also known as
Q1157+014 or PKS1157+014, is a radio-loud quasar \citep{Wall71}.  
\cite{Ganguly13} found that radio-loud quasars tend to have intrinsic absorption systems 
compared to radio-quiet ones, although the statistics for the former could be greatly affected by a 
single extraordinary object.
  \cite{Hayashi13} placed a constraint on a viewing angle of the
  quasar smaller than $73-77^{\circ}$ with respect to the polar-jet
  based on radio polarimetric observations, while no constraint is
  placed on the lower limit. The quasar is not pole-on viewed at
  least, since it shows two-sided radio structures without significant
  polarization in its central component.

\begin{figure}
	\includegraphics[width=\columnwidth]{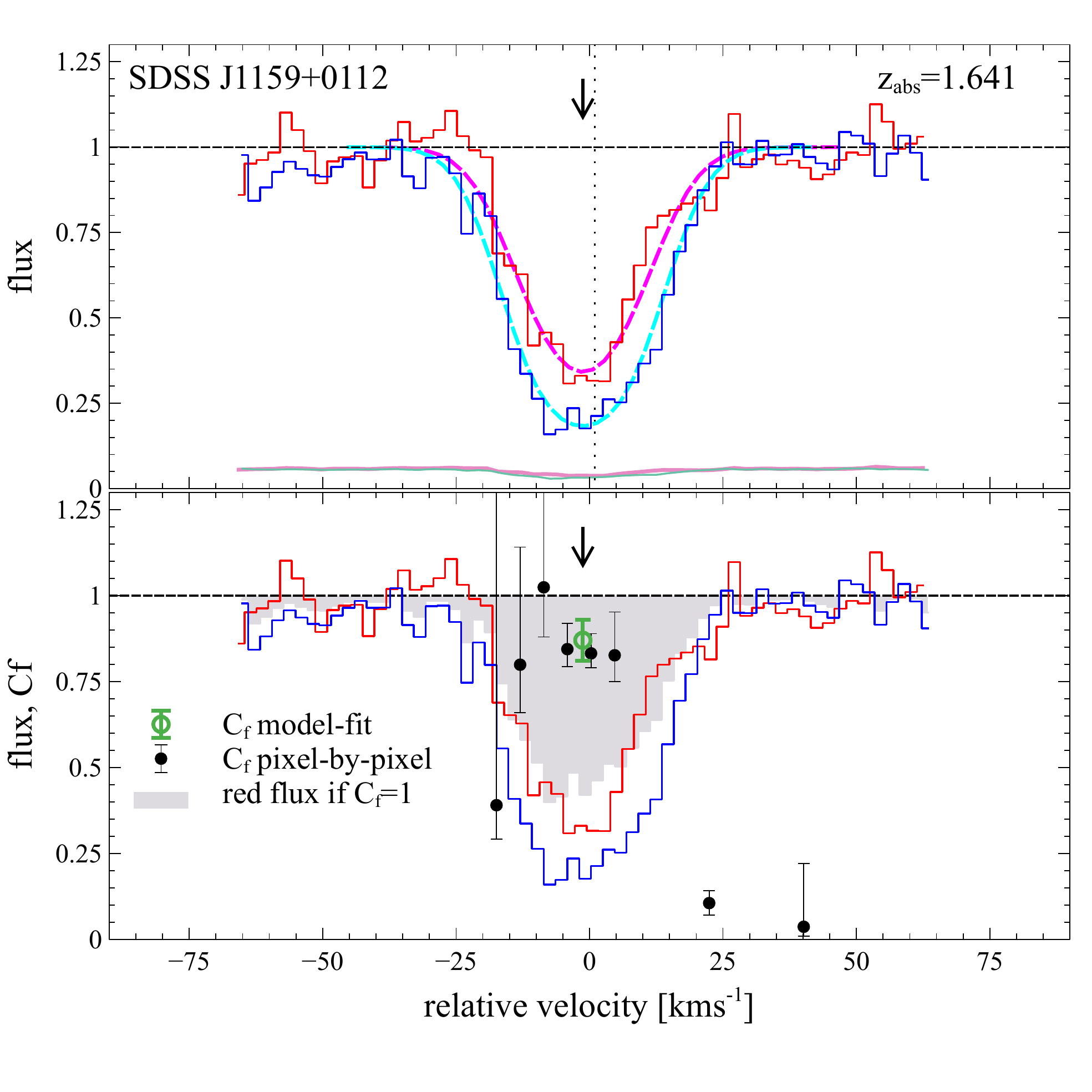}	
	\caption{Same as Figure~\ref{fig:J02421945}, but for
          \ion{C}{4} NAL at $z_{\rm abs}=1.641$ in
          J1159+0112.}
	\label{fig:J11591641}
\end{figure}

J1159+0112 has very strong \ion{C}{4} and \ion{N}{5} BALs 
 in the spectrum with FWHM of $> 3,000$~\kms around emission
line peaks. In addition, there exists a mini-BAL system at $v_{\rm
  off} \sim 8,000$~\kms. \citet{Misawa14} monitored the mini-BALs and
detected an obvious time variability. The spectrum we use was taken
with VLT/UVES in June 2007, covering a wavelength range of $\sim
3,700-5,000$~\AA\ and $\sim 5,600-9,600$~\AA. The \ion{C}{4} NAL at
$z_{\rm abs} =1.641$ is detected at wavelength region whose S/N ratio
is large enough (S/N $\sim$18 pix$^{-1}$) for a partial coverage
analysis. Best-fit model for the \ion{C}{4} NAL in
Fig~\ref{fig:J11591641} provides a partial coverage of
$C_{\mathrm{f}}=0.87 \pm 0.06$. Since both the fitting and the 
pixel-by-pixel methods support that a covering factor is smaller than
unity with a confidence level of $> 2\sigma$, we classify the system
into class~B.

Our spectrum covers only wavelength regions corresponding to
\ion{Si}{4} and \ion{Mg}{2} lines other than \ion{C}{4}, however, the
former is severely blending with strong \ion{N}{5} BAL and the latter
is not clearly detected. Therefore, it is impossible to estimate the
ionization condition of the NAL system.

This system has a quite large ejection velocity with $v_{\rm
  ej}\sim38,000$~\kms, much larger than those of BAL and mini-BAL
systems. Same as the case of J0242+0049, such a large offset velocity
implies that this system is strongly accelerated by powerful
quasar-driven outflow winds.

\subsubsection{Class~B NAL at $z_{\rm abs}=2.691$ in  J1215-0034}
The quasar J1215-0034 ($z_{\rm em} = 2.71$, $m_{\rm g} =17.50$), also
known as Q1213-003, has the strongest \ion{C}{4} (and also detected
\ion{Si}{4}, \ion{N}{5} and \lya) BAL with $\mathrm{BI_{0}}=
4807.7$~\kms\ among our sample.  This quasar was regarded as HiBAL
quasar in \cite{Bruni19}.  High resolution spectra ($R$ $\sim$ 51690)
of the quasar were taken with VLT/UVES from March 28 to April 1 in
2011, covering a wavelength of $\sim 3,300-4,500$~\AA\ and $\sim
4,600-6,800$~\AA.

\begin{figure}
	\includegraphics[width=\columnwidth]{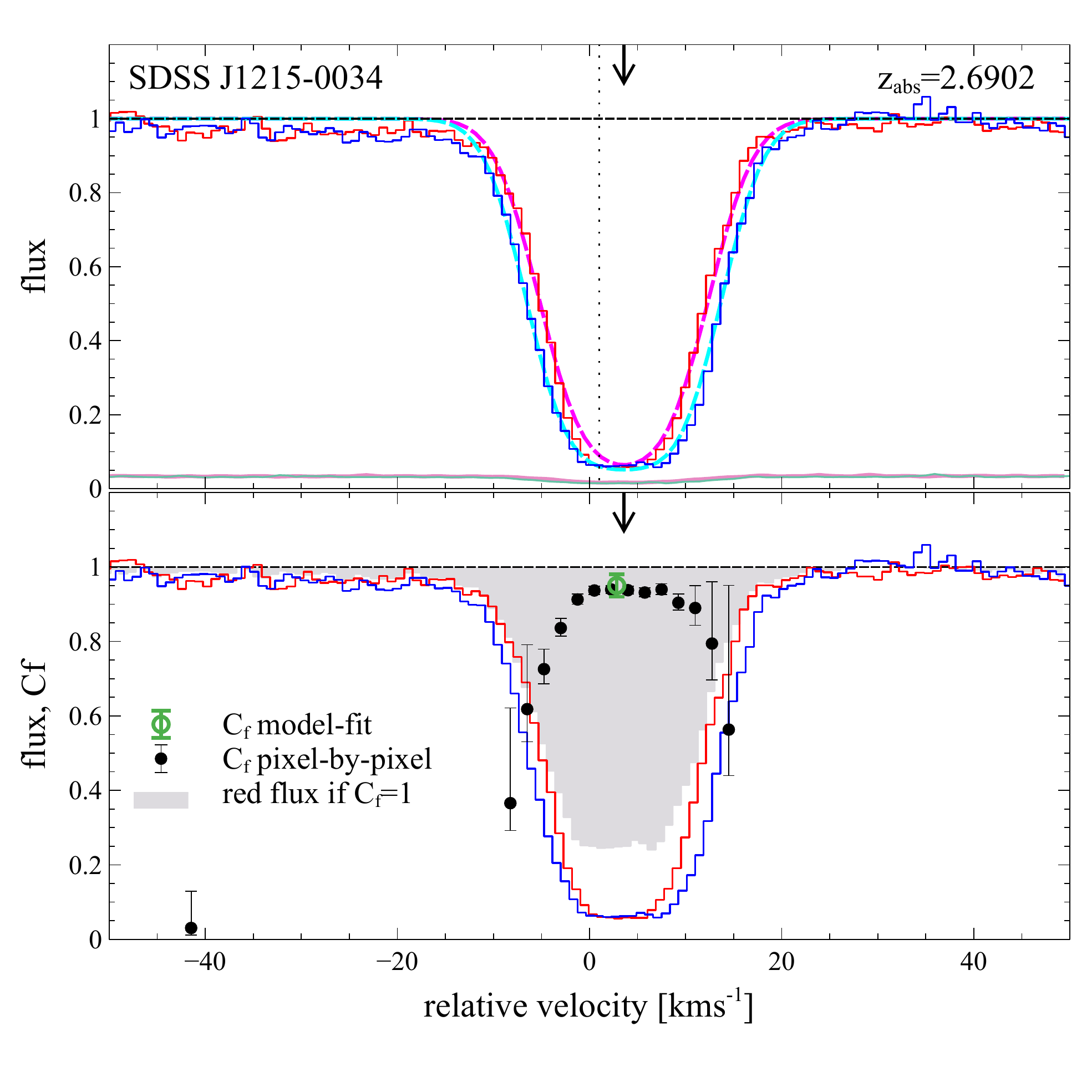}	
	\caption{Same as Figure~\ref{fig:J02421945}, but for
          \ion{C}{4} NAL at $z_{\rm abs}=2.691$ in
           J1215-0034.}
	\label{fig:J12152691}
\end{figure}

\begin{figure*}
	\includegraphics[width=180mm]{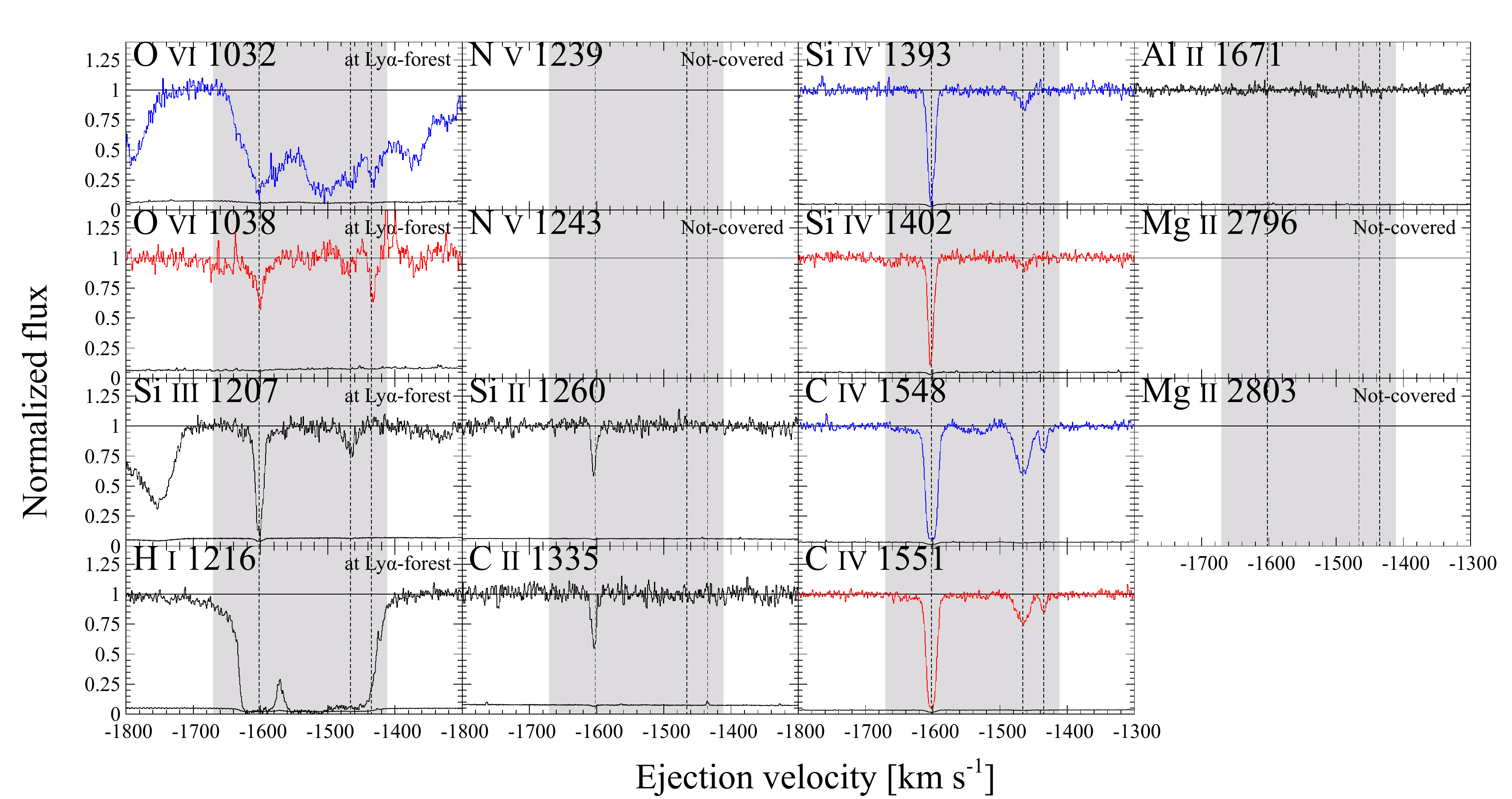}	
	\caption{Same as Figure~\ref{fig:J02421945sys2}, but for NAL
          system at $z_{\rm abs}=2.691$ in J1215-0034.  Shaded gray
          regions show the line width of the broadest absorption line
          without line blending in the system (i.e., \lya), where a
          normalized flux remains below 0.9.}
	\label{fig:J12152691sys2}
\end{figure*}

A \ion{C}{4} NAL at $z_{\rm abs} = 2.691$ is highly likely associated
to the background quasar, the host galaxy, or its surrounding environments 
because its offsetvelocity from the quasar is smaller than $5,000$~\kms\ (i.e., an
associated absorption line (AAL) as defined by \citealt{Weymann79} and
\citealt{Foltz86}).
Since the \ion{C}{4} NAL is detected in the spectral region whose S/N
ratio is quite high ($\sim 27$ pix$^{-1}$), our result of the covering
factor analysis is very reliable. The result of our fitting model to
the \ion{C}{4} NAL is shown in Fig.~\ref{fig:J12152691}, giving
covering factor of $C_{\rm f}$ = $0.95\pm0.03$ at the center. Since
this result is consistent to the pixel-by-pixel method, we classify
the NAL into class~B. Our result is also visually inspected; both blue
and red members of \ion{C}{4} doublet have almost the same absorption
depth at the center with some residual flux (see
Fig.~\ref{fig:J12152691}), which is usually called as ``non-black
saturation'' and one of the most reliable signature of partial
coverage.  On the other hand, \ion{Si}{4} NALs in the same system are
not showing partial coverage.  However, this is not so surprising
because partial coverage is often seen only in a part of transitions
in same systems \citep{Misawa07}.  Since this system shows various
ionic absorption lines from low-ions (e.g., \ion{C}{2} and
\ion{Si}{2}) to high-ions (e.g., \ion{O}{6}) as shown in
Fig. \ref{fig:J12152691sys2}, it may have a complex ionization
gradient inside it. Strong {\lya} absorption line in this system has a
large line width of FWHM $\sim$~200~\kms\ and almost saturated
throughout the profile.

\section{Discussion}

\subsection{Intrinsic NAL Absorbers along Sight-lines toward BAL Quasars}

  Our detection of intrinsic (i.e., class A/B) NALs in BAL/mini-BAL
  quasar spectra suggests that intrinsic NAL absorbers locate along
  our LOS not only to non-BAL quasars but BAL/mini-BAL quasars. The
  detection rates of intrinsic NALs toward both types of quasars are
  roughly in agreement (i.e., $\sim 30$\%), but it should be regarded
  with some cautions because i) there is a large statistical
  uncertainty of our results for BAL/mini-BAL quasars with a detection
  rate of $33^{+33}_{-18}\%$ due to a small sample size, and ii) we
  placed only a lower limit on the detection rate since some fraction
  of intrinsic NALs could have a full coverage.

  \cite{Misawa07} found intrinsic \ion{C}{4} NALs in
  $\sim32$\%\footnote{The fraction would be $\sim50\%$ if we also
    consider \ion{N}{5} and \ion{Si}{4} in addition to \ion{C}{4}. }of
  non-BAL quasars at $z\sim2-4$.  \citet{Simon12} also obtained a
similar fraction for \ion{C}{4} NALs ($\sim37$\%) in a velocity range
of $v_{\rm off} \sim 2,500-40,000$~\kms. These results were obtained
based on a partial coverage analysis as we adopted. Using time
variability analysis, \citet{Wise04} discovered that 4 out of 15
($\sim27$\%) quasars have at least one intrinsic \ion{C}{4} NAL at
$v_{\rm off} < 5,000$~\kms. \citet{Nestor08} had a slightly small
value for the fraction ($\sim14$\%) at $v_{\rm off} < 12,000$~\kms,
but their result is based on a number excess of \ion{C}{4} NALs
without identifying intrinsic NALs using low-resolution SDSS spectra.
For further discussion to compare the detection rate of intrinsic
  NALs between BAL and non-BAL quasars, we need to increase (at least
  triple) our sample size that is comparable to the sample size of
  \cite{Misawa07}.

In the unified model of outflows, intrinsic NALs tend to be observed
at small inclination angle relative to the rotational axis of the
accretion disk, while BALs and mini-BALs are observed when our LOS
passes through the outflow wind at larger inclination angle (i.e.,
closer to the accretion disk).  In our analysis, however, we detect
intrinsic NALs in spectra of $\sim33^{+33}_{-18}$\%\ of BAL quasars,
which suggests that the location of intrinsic NAL absorbers are not
limited to the regions at higher latitude but they exist everywhere
regardless of inclination angle. Thus, the simple unified model of
outflow wind is not applicable.

\subsection{Radial Distance of NAL system at $z_{\rm abs}=2.691$ in  J1215-0034}

As shown in Fig. \ref{fig:J12152691sys2}, we detected \ion{Si}{2} and
\ion{C}{2} lines in the NAL system at $z_{\rm abs}=2.691$ in
J1215-0034.  Since Si$^+$ and C$^+$ ions have three electrons at
outermost electron shell, the ground electron state has two
fine-structure levels (true ground $J=1/2$ and slightly excited
$J=3/2$).  As done in the literature (e.g., \citealt{Barlow97};
\citealt{Hamann01}; \citealt{Borguet12}; \citealt{Chen18}), it is
possible to estimate a gas density and a radial distance of absorbers
from the central source using photoionization model for absorption
systems with both ground and excited levels such as
{\ion{S}{4}}$^{*}$/\ion{S}{4}, \ion{Si}{2}$^{*}$/\ion{Si}{2},
\ion{C}{2}$^{*}$/\ion{C}{2} and \ion{Fe}{2}$^{*}$/\ion{Fe}{2}.  An
ionization parameter $U$, the ratio of hydrogen-ionizing photons
illuminated to absorber ($n_{\mathrm{\gamma}}$) and the total hydrogen
number density ($n_{\mathrm{H}}$) is defined as
\noindent 
\begin{equation}
\label{eq2}
 U\equiv \frac{n_{\mathrm{\gamma}}}{n_{\mathrm{H}}}=\frac{Q({\mathrm H})}{4{\pi}cn_{\mathrm{H}}R^2},
\end{equation}
where $Q({\mathrm H})$ is the hydrogen-ionizing photon number rate emitted by
central source, and $R$ is a distance of absorbers from the central
source.

To measure a radial distance $R$, we also need to know
$n_{\mathrm{H}}$.  In highly ionized zone, the relation between the
$n_{\mathrm{H}}$ and an electron density $n_{\mathrm{e}}$ is
approximately shown as $n_{\mathrm{e}}\sim1.2n_{\mathrm{H}}$.  The
$n_{\mathrm{e}}$ can be estimated using a column density ratio of ground
and excited levels by

\noindent 
\begin{equation}
\label{eq3}
 n_{\mathrm{e}}=n_{\mathrm{cr}}{\Big[\frac{N_l}{N_{u}}\bigg(\frac{g_{u}}{g_{l}}\bigg)e^{-\Delta E/{kT}}-1\Big]}^{-1},
\end{equation}
where $N_l$ and $N_u$ are column densities of ground and excited
levels, ${g_u}~/~{g_l} = 2 $ is a ratio of statistical weight, $\Delta
E$ is the energy difference between the two-levels (e.g., $63.42~\cm$
for \ion{C}{2}$^{*}$ and \ion{C}{2} from the NIST Atomic Spectra
Database
  \footnote{ https://physics.nist.gov/asd }),
$k$ is Boltzmann constant, $T$ is a nominal temperature of
ionized plasma, and $n_{\rm cr}$ is the critical density for excited and
ground state ($\sim50~\cmmm$ at $T\sim10^4~\K$ for \ion{C}{2}$^{*}$/\ion{C}{2}; 
e.g., \citealt{Goldsmith12}).  

\begin{figure}
	\includegraphics[width=\columnwidth]{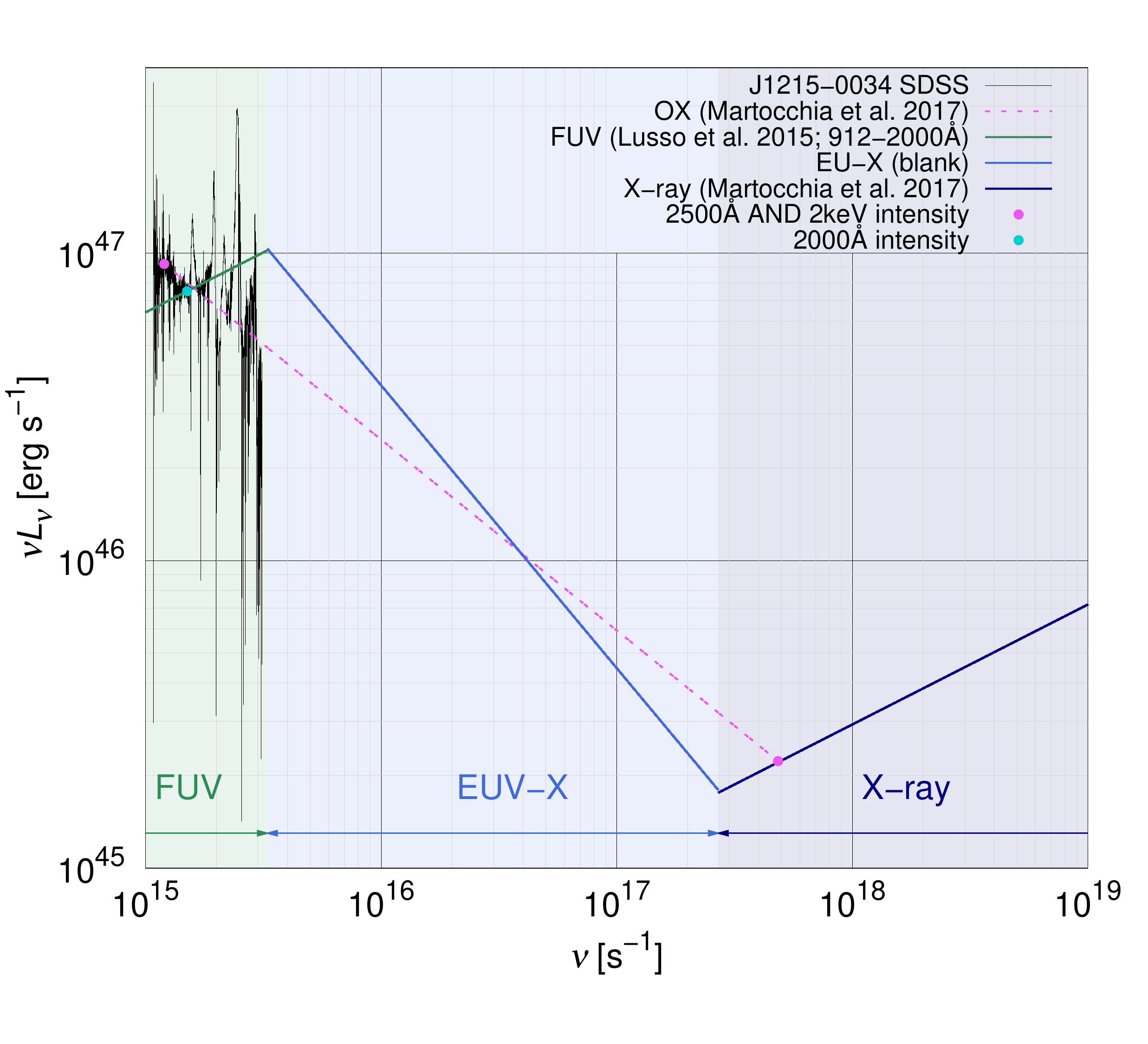}	
	\caption{The SED model of J1215-0034. We plotted SDSS spectrum
          of J1215-0034 in black solid line.  We used at 2000~\AA {}
          intensity (cyan circle) for normalizing SED models.  Magenta
          circle and dot line represent 2500\AA $~$and 2keV
          monochromatic luminosity ratio (using result of
          \citealt{Martocchia17}). The solid lines of green, blue and
          dark-blue represent SED models of FUV, EUV-X and X-ray
          regions, respectively. }
	\label{fig:SED}
\end{figure}

We assume an ionization parameter of $\log~U\sim -2.0$ based on the
ratio of column density of \ion{C}{2}$/$\ion{C}{4} and
\ion{Si}{2}$/$\ion{Si}{4} from \textsc{cloudy} models in
\cite{Hama97}.  We also need SED models for estimating an ionizing photon
rate $Q(\mathrm{H})$.  The MF87 SED (\citealt{MF87}) has been widely
used as a typical AGN SED model.  However, as discussed in
\cite{Dunn10} and \cite{Arav13}, the MF87 model could be only
applicable for radio-loud quasars.  For radio-quiet quasars, it is
recommended to use UV-soft SED model (i.e., not showing UV-bump around
600~\AA). Because J1215-0034 is a radio-quiet quasar with a
radio-loudness of ${\mathcal R}<~1.76$ (see
Table~\ref{tab:SDSSsample}), we use a UV-soft SED below.  From near-UV
(NUV) to far-UV (FUV) region (i.e., $912-3,000$~\AA), we adopt
spectral index $\alpha_{\mathrm{UV}}= - 0.61$ from \cite{Lusso15} and
scale it to UV flux of J1215-0034.  \cite{Lusso15} have determined an
average spectral index from NUV, FUV, to extreme-UV (EUV), for
high-redshift ($2.3<z_{\rm em}<2.6$) and bright ($g < 18.5$~mag)
quasars. The quasar J1215-0034 ($z_{\rm em} = 2.71$ and $g =
17.50$~mag) roughly satisfy the criteria above.  In X-ray, we adopt
spectral index $\alpha_{\mathrm{X}} = -0.61$ and optical-to-X-ray
slope $\alpha_{\mathrm{OX}} = -1.62$ from observations with $Chandra$
(\citealt{Martocchia17}).  Finally, we interpolate a power law between
FUV and X-ray to estimate an SED in EUV which is observationally
unavailable.  We show the estimated SED model in Fig.\ref{fig:SED}.
As a result, we calculate $Q(\mathrm{H})=2.43\times10^{57}~
{\mathrm{s}^{-1}}$.

\begin{figure*}
	\includegraphics[width=140mm]{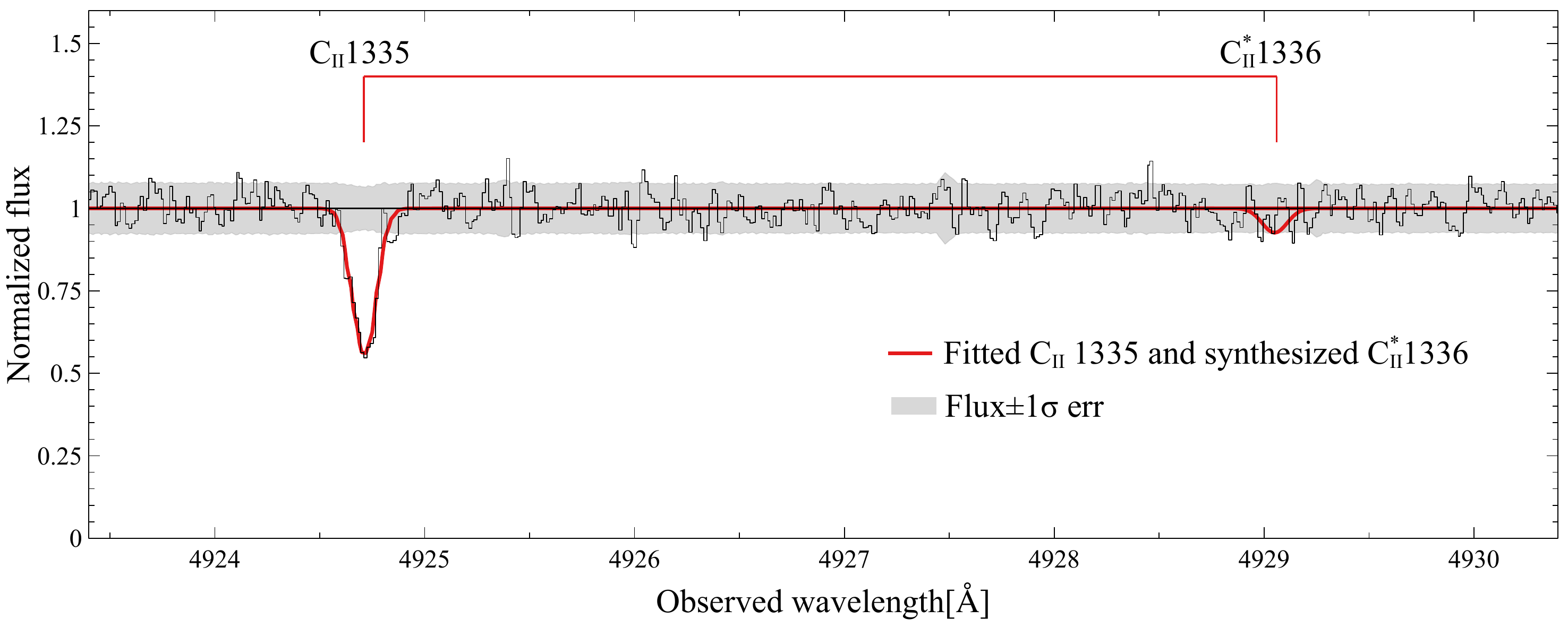}	
	\caption{Observed (black histogram) and modeled (red curves)
          profiles of ground \ion{C}{2} 1335 and excited
          \ion{C}{2}$^{*}$ 1336 lines in a spectrum of J1215-0034.
          The shaded gray region show $1\sigma$ uncertainly of
          continuum flux.}
	\label{fig:CII1335-6}
\end{figure*}

\begin{figure}
	\includegraphics[width=\columnwidth]{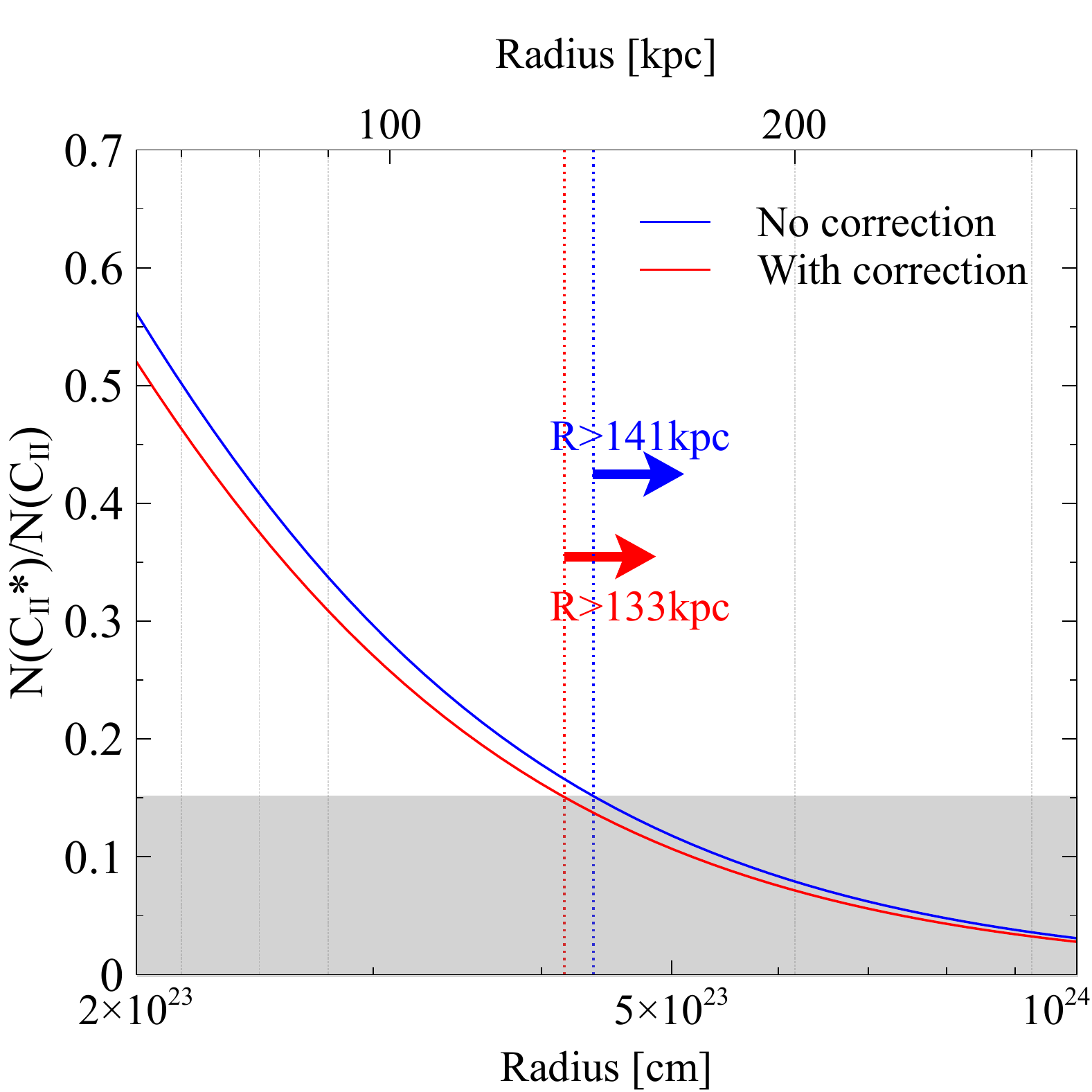}	
	\caption{Line ratio of \ion{C}{2}$^*$ and \ion{C}{2} as a
          function of radial distance for the intrinsic NAL absorber
          in J1215-0034. The blue and red curves show the relations
          without and with BAL absorption correction.  The shaded
          region in gray indicates N(\ion{C}{2}$^{*}$){ }$/${
          }N(\ion{C}{2})${ }<0.15$, corresponding to a radial distance
          of $R> 141~\kpc$ (blue dotted line) and $R>133~\kpc$ after
          the correction of the BAL attenuation (red dotted line),
          respectively.}
	\label{fig:location}
\end{figure}

We detect ground levels of \ion{Si}{2} and \ion{C}{2} lines in the NAL
system, but no remarkable profiles are detected for excited levels.
Therefore, it is possible to place an upper limit on $n_{\mathrm{e}}$
and a lower limit on $R$, using equations \ref{eq2} and \ref{eq3}.
Because $n_{\rm cr}$ of \ion{C}{2} is smaller than that of
\ion{Si}{2}, we calculate these limits only for \ion{C}{2} line to
place more stringent constraints on $n_{\mathrm{e}}$ and $R$.  While
we estimate a column density of ground level of \ion{C}{2} line as
$\log{N_l} (\cmm)=13.05$, we place only upper limit on it for
\ion{C}{2}$^*$ as $\log{N_u} (\cmm) \leq 12.23$ assuming 1$\sigma$
uncertainty of continuum level.  Fig. \ref{fig:CII1335-6} shows
modeled profiles of ground levels of \ion{C}{2} line. The ratio of
column density between ground and excited levels is $N_{
  u}~/~N_{ l} \leq 0.15$, which leads to an upper limit on the
electron density ($n_{\rm e} \leq 4.1~{\cmmm}$), and a lower limit on
the distance of absorber is $R \geq141~{\kpc}$, as shown in
Fig. \ref{fig:location}.

Possible concern about this calculation is how much an incident
ionizing flux from the continuum source is dimmed by BAL
absorber. Because the NAL system locates along our LOS to the quasar
behind the BAL absorber, we could overestimate ionizing photons to the
NAL absorber.  If we simply assume that ionizing photons are partially
absorbed only by neutral hydrogen in the BAL absorber, an optical
depth at the Lyman limit is
\begin{equation}
\label{eq4}
\tau_{LL}=N_{\mathrm{HI}}\sigma^{0}_{\mathrm{H}},
\end{equation}
where $\sigma^{0}_{\mathrm{H}}=6.3\times10^{-18} {\CMM}$ is the
\ion{H}{1} cross section at the Lyman limit and $N_{\mathrm{HI}}$ is
\ion{H}{1} column density of the BAL system.  We evaluate
$N_{\mathrm{HI}}$ applying apparent optical depth method (i.e., AOD,
\citealt{Savage1991}) to the {\lya} BAL. Using equations (10) and (11)
from \cite{Savage1991}, we estimate $\log N_{\mathrm{HI}}
(\cmm)=16.10$, and a fraction of ionizing photons that are attenuated
by the BAL absorber is only $\sim$10\%. Even if we consider this
correction, a lower limit of the radial distance is still large ($R
\geq 133~\kpc$). Therefore, we conclude that the NAL system locates at
a large distance comparable to a scale of Circum-galactic medium
(CGM)\footnote{This is not a scale of the IGM since the size of CGM is
  usually defined as a virial radius $r_{\mathrm{vir}}$, which is
  $\sim200~\kpc$ for $L^{*}$ galaxies \citep{Tumlinson17}.}  which is
several orders of magnitude larger than the typical scale of
radiation-driven wind at $\sim0.01-0.1~\pc$ (\citealt{Murray95},
\citealt{Proga00}).

\subsection{Two types of intrinsic NALs}

\subsubsection{Intrinsic NAL in  J1215-0034}

This NAL system is a typical associate absorption line (AAL) because
it has a small ejection velocity ($v_{\rm off} \sim
1,600$~\kms). Possible origins of AAL absorbers include i) AGN outflow
winds in pc to sub-pc scale, ii) materials associated with quasar host
galaxies, and iii) cluster/group of galaxies around of the quasar.
Among these, \cite{Wild08} concluded that about 40\% of \ion{C}{4}
AALs with $v_{\rm off} \leq 3,000$~\kms\ are probably physically
associated with the quasars themselves and its host galaxies [i.e.,
  case i) and ii) above].
If we assume \ion{C}{4} broad emission line region (BELR) is a
background source, the size of NAL absorber is smaller than the size
of BERL $R_{\mathrm{CIV}}=0.12\pm0.10~\pc$, where we use an
equation~(1) in \cite{Lira18}.  This size is much smaller than the
size of typical cosmologically intervening absorbers (i.e.,
$1-100~\kpc$).  Thus, among the three possible origins the case
  iii) is less likely since the corresponding size of absorbers
  ($\sim$kpc) is much larger than the size of BELR.

Among the other possible origins, the case i) is also rejected because
the absorber's radial distance from the central source ($R>130~\kpc$)
is much larger than the typical radial distance of radiation-driven
outflow wind (i.e., $0.01-0.1~\pc$).  Thus, our results suggest the
NAL system most likely originates in materials associated with quasar host
galaxies or CGM around it.

Indeed, recent results of radial distance measurements of BAL/mini-BAL
absorbers with photoionization models using ground/excited state lines
implies that they sometimes spread up to $0.1-10~\kpc$
(\citealt{Moe09}; \citealt{Dunn10}; \citealt{Aoki11};
\citealt{Borguet13}; \citealt{Chamberlain15}; \citealt{Arav18};
\citealt{Xu18} )  or $>10~\kpc$ (\citealt{Hamann01};
  \citealt{Hutsemekers04}; \citealt{Borguet12}) from the central
  source.  \cite{Faucher12} proposed a model that BAL absorbers 
  especially for LoBAL ones form in situ in the ISM of the host
galaxy.   For intrinsic NALs, \cite{Wu10} estimated their radial
  distance as $0.01-10~\kpc$ from the central source based on
  photoionization models.  Considering these results, intrinsic NAL
absorbers at a large distance with $R>100~\kpc$ may have some
relationship with such kilo-parsec scale BAL/NAL absorbers. 
They could be decelerated while interacting with materials in/around 
host galaxies, and eventually contributing to energy/momentum feedback 
to host galaxies, CGM, and IGM around them.

  The NAL system has other possible origins since there exist a
  point source $\sim 5 $ arcsec (corresponding to$\sim40~\kpc$ at
  $z\sim2.69$) north-west of J1215-0034 both in {\it Chandra} X-ray
  and SDSS DR16 optical images. If this is a foreground faint AGN at
  $z\sim 2.69$, the NAL system could be due to outflow wind from the AGN
  whose opening angle is large enough to cover our LOS to
  J1215-0034. Alternatively, if this point source is a normal galaxy
  (i.e., inactive galaxy), its CGM could produce the NAL at
  $z\sim2.69$. If the latter is the case, the corresponding absorber 
  should be in a special condition because its size is small enough not to
  cover the background flux source of J1215-0034 despite being an intervening 
  absorber.

  Recently, \cite{Balashev20} discovered a proximate Damped
  \lya~system (PDLA) with a partial coverage at a distance of $\sim
  150-200~ \kpc$ from the background quasar. They infer the PDLA
  originates from a galaxy located in a group where the quasar-host
  galaxy resides. The properties (i.e., size and radial distance) of
  this PDLA is very similar to those of the class B NAL in
  J1215-0034. If this NAL system originates in an absorber like a PDLA above, 
  some fraction of intrinsic NAL systems in the past studies could also
  have similar origins (i.e., absorbers in member galaxies in
  group/cluster around quasar host galaxies). In the current
  study, we conclude that the NAL system at $z\sim 2.69$ is physically
  associated to J1215-0034 (i.e., intrinsic NAL) based on our already
  available data, however, we will conduct spectroscopy of the point
  source ($m_{\mathrm g} \sim 21.3$) to locate its origin as a future work.

\subsubsection{Intrinsic NALs in SDSS J024221.87$+$004912.6 and J1159+0112}
For the other two intrinsic NAL systems, we cannot place strong
constraints on their radial distance without detecting either time
variability or absorption lines from fine-structure levels.  However,
we have several important properties of them.

The NAL system in J0242$+$0049 shows three signs of intrinsic origin:
a) only high-ionization absorption lines such as \ion{N}{5},
\ion{C}{4} (and possible \ion{Si}{4}) are detected without any
remarkable detection of low-ionization lines as shown in Fig
\ref{fig:J02421945sys2}, b) it has very large ejection velocity of
$\sim11,600$~\kms with relatively broad ($\sigma_v > 100$~\kms) and
smooth line profile, and c) it shows a reliable (i.e., class A)
partial coverage along LOS to the continuum source with size of
$\sim0.01~\pc$.  These properties strongly support that the NAL system
indeed originates in a small clumpy (or filamentary) structure in
central region of the outflow wind, and being accelerated by
radiation-pressure. Such a small scale clumpy structure is frequently
reproduced by two/three-dimensional radiation-magnetohydrodynamic
simulations caused by hydromechanic instability in the innermost
region of outflow wind (\citealt{Ohsuga05}, \citealt{Takeuchi13}, and
\citealt{Kobayashi18}).  Although only \ion{C}{4} line is detected in
another NAL system in J1159+0112, it also has some properties of
intrinsic NALs like partial coverage and large ejection velocity
($\sim38,000$~\kms). It could be in a same environment as the NAL
system above but being accelerated more efficiently.

\subsection{NAL/BAL absorbers at large and small radial distances}

In the previous section, we discuss two extreme types of intrinsic NAL
absorbers: a high-ionized NAL with large ejection velocity in
J0242+0049 and a low-ionized NAL with small ejection velocity in
J1215-0034 (we cannot classify the NAL in J1159+0112 because only
\ion{C}{4} is detected). The former is probably related to the outflow
wind that is being radiatively accelerated at a small distance from
the continuum source, while the latter locates at a large distance
comparable to the scale of CGMs.  Thus, these NAL systems may
correspond to different types of absorbers.

Indeed, such a large difference in a radial distance has been
suggested for BAL absorbers repeatedly.  As a traditional method, a
variability time scale of BAL strength/profile has been used for
calculating a radial distance of BAL absorbers assuming a Keplerian
motion.  This method tends to give small distances (e.g., $R<10~\pc$;
\citealt{Capellupo11}; \citealt{Rodriguez11}; \citealt{Muzahid16};
\citealt{Moravec17}; \citealt{McGraw17}).

Another method of calculating absorber's distance based on
photoionization models would lead to larger values. For example, a
distance of $\sim 1-10~\kpc$ was evaluated by comparing line strengths
of ground/excited levels of relatively low-ionized ions (\ion{Si}{2},
\ion{C}{2} and \ion{Fe}{2}) with a small ejection velocity $\sim
5,000$~\kms\ (\citealt{Moe09}; \citealt{Dunn10}; \citealt{Aoki11}).
If higher ionized ions like \ion{S}{4} with larger ejection velocity
of $\sim 10,000$~\kms are used, the value would be somewhat smaller,
$\sim0.1~\kpc$ (\citealt{Borguet13}; \citealt{Chamberlain15};
\citealt{Arav18}; \citealt{Xu18}).

Here, we would like to emphasize a measured radial distance depends on
calculating methods. For example, we tend to derived a small radial
distance based on time variability analysis assuming a gas motion 
since outflow winds which locate in the vicinity of central source are rotating 
much faster than those at large radial distance ($>~1~\kpc$).  This trend is consistent
to the past results that BALs with large ejection velocity (i.e.,
rapidly rotating around the central source) tend to show time
variability in smaller time scales compared to those with smaller
ejection velocity (\citealt{Capellupo11}).

\subsection{Possible geometry of BAL and intrinsic NAL absorbers}

We conclude that an inclination angle alone cannot be a good indicator to 
distinguish intrinsic NAL from BAL absorbers as frequently
introduced in the literature, because 1) intrinsic NALs are detected 
regardless of an existence of BAL/mini-BAL features in quasar spectra,
and 2) both BAL and intrinsic NAL absorbers distribute over a wide range 
of radial distance from the central source (\S~5.4).  

\begin{figure}
	\includegraphics[width=\columnwidth]{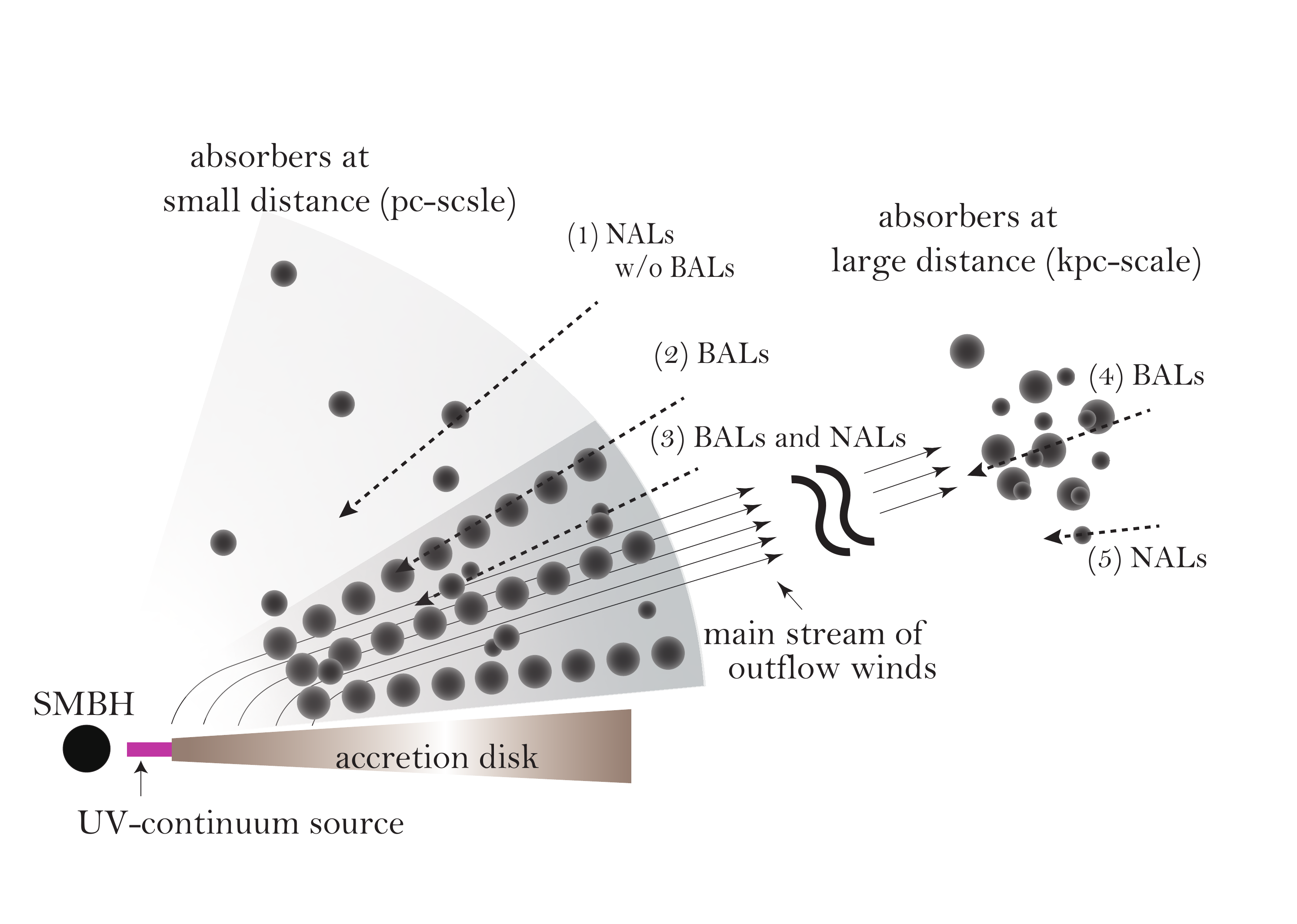}	
	\caption{Possible geometry of outflowing absorbers. At smaller
          distances, there exist small clumpy clouds (black sphere)
          that can be observed as intrinsic NALs. The main stream of
          outflow wind consisting of a large number of clumpy clouds
          would be observed as BALs. At larger distances, clumpy
          clouds are formed by interaction between outflow gas and
          ISM/CGM of host galaxy and they are observed as BALs/NALs
          depending on how many clouds are clustering. Dashed line (1)
          represents a LOS along which we observe intrinsic NALs at
          small inclination angles.  Dashed lines (2) and (3) are LOS for
          BALs, but only in line (3), clumpy clouds with different
          offset velocity from BAL absorber are observed as intrinsic
          NALs in addition.  At large distances, absorption line width
          also depends on a number of clumpy clouds as shown with two
          dash lines (4) and (5).}
          	\label{fig:incli}
\end{figure}

We present a possible structure of outflow winds based on our results
in Fig. \ref{fig:incli}, in which there exist both small distance
absorbers at pc-scale and large distance ones at
$\kpc$-scale or more. The size of intrinsic NAL absorbers should be
smaller than the central flux source because of partial coverage.  The
small distance absorbers are probably related to the radiation-driven
outflows near the center and accelerated up to $\sim 0.1${\it c},
depending on the acceleration efficiency and/or an inclination angle.
On the other hands, the large distance outflows are probably extended
outflow winds \citep[e.g.,][]{Chen18} that are formed (and
decelerated) by interaction with ISM and CGM of host galaxies. In this
model, we are able to observe intrinsic NALs in any inclination
angles.

It could be possible that some fraction of BALs and intrinsic NALs have
same origins but their line width depends on a number of small scale
clumpy clouds (i.e., larger number of clouds would produce broader
absorption profile like BALs).  \cite{Lu18a, Lu18b, Lu19} attempted to
classify BALs into two subclasses: Type~S BALs (BALs with smooth
trough) and Type~N BALs (BALs with multiple components like NALs), and
suggested that the latter is essentially originated in same absorbers
as intrinsic NALs. Our results supports the scenario. 

\section{Conclusion}

In this study, we search for intrinsic NALs in 11 BAL/mini-BAL quasar
spectra using partial coverage analysis, to test whether intrinsic
NALs are detected along with BALs/mini-BALs.  We summarized our
results below:

\begin{enumerate}

\item We identified one reliable (class~A) and two possible (class~B)
  intrinsic NAL system in 3 of 9 quasars with clear BAL profiles, which
  suggests that the location of intrinsic NAL absorbers are not
  limited to the regions at higher latitude and that they exist
  everywhere regardless of inclination angle.

\item Using photoionization model with ground/excited levels of
  \ion{C}{2} absorption line, we estimated a radial distance of
  intrinsic NAL absorber in J1215-0034 as $R>140~\kpc$ that is
  comparable to the scale of the CGM of host galaxy. A lower limit of
  radial distance is still large ($R>130~\kpc$) even after considering flux dimming
  by the BAL absorber along the same LOS.

\item The intrinsic NAL system in J0242+0049 showing possible partial
  coverage may originate in small clumpy clouds at small radial
  distance because it has a large ejection velocity and high
  ionization condition, while the NAL system in J1215-0034 has
  opposite properties (i.e., small ejection velocity and low
  ionization condition).

\item Intrinsic NAL systems in J1215-0034 and J0242+0049 with
  different physical properties may arise at different type of
  absorbers at very different radial distances, as sometimes suggested
  for BAL absorbers. The former could be the outflow winds at large
  distance that are interacting with ISM and CGM of host galaxies.
  
\item BAL and intrinsic NAL absorbers could be closely related each
  other because both class of absorbers locate at wide range of radial
  distance along same LOS.  Thus, our results suggest that absorption
  width does not simply depends on the inclination angle, but also
  depends on internal structure of outflow winds.

  \vspace{10mm}
  
    As a future work, we will increase our sample
    size at least by a factor of three to match the sample size for
    non-BAL quasars (37 in \citealt{Misawa07}) to investigate how the
    detection rate of intrinsic NALs depends on the co-existence of
    BAL features along same LOS.  We also need to spectroscopically identify the point
    source at $\sim 5$ arcsec north-west of J1215-0034 to narrow down possible origins of the class B
    NAL absorber with a quite large radial distance of $\geq130~\kpc$  (i.e., a large distance outflow 
    or an unusual intervening absorber with a partial coverage  in a foreground galaxy).
  
 \end{enumerate}

\section*{Acknowledgements}
The research was supported by the Japan Society for the Promotion of
Science through Grants-in-Aid for Scientific Research 18K03698.

Based on observations made with ESO Telescopes at the La Silla Paranal
Observatory under programs, 081.A-0479(A), 075.B-01908A9,
083.A-0042(A), 092.B-0574(A),079.B-0479(A), 273.A-5020(A),
080.A-0482(A), 185.A-0745(D), 081.A-0334(A), 091.A-0018(A),
081.B-0285(A).

\section*{DATA AVAILABILITY}
The datasets underlying this article were derived from sources in
the public domain as given in the respective footnotes.












\bsp	
\label{lastpage}
\end{document}